\documentclass{conm-p-l}

\usepackage{amssymb,epsfig}
\numberwithin{equation}{section}

\allowdisplaybreaks
\newcommand{\HALF}{{\textstyle\frac{1}{2}}}

\newcommand{\TFRAC}[2]{{\textstyle\frac{{#1}}{{#2}}}}

\newcommand{\OL}{\overline}
\newcommand{\UL}{\underline}

\newcommand{\DAB}[1]{\rule[0pt]{0pt}{#1}}

\newcommand{\EMB}{\boldsymbol}
\newcommand{\VEC}[1]{{\EMB{#1}}}
\newcommand{\LAB}[1]{{\mathsf{#1}}}
\newcommand{\FUN}[1]{{\mathrm{#1}}}
\newcommand{\MAT}[1]{{\mathbf{#1}}}
\newcommand{\TEN}[1]{{\mathfrak{#1}}}
\newcommand{\GRP}[1]{{\mathsf{#1}}}
\newcommand{\FLD}[1]{{\mathbb{#1}}}
\newcommand{\ALG}[1]{{\mathcal{#1}}}

\newcommand{\TR}{\FUN{tr}}

\newcommand{\UPS}{\VEC\iota}

\newcommand{\RHO}{{\EMB\rho}}
\newcommand{\SIG}{{\EMB\sigma\!}}
\newcommand{\GAM}{{\EMB\gamma}}
\newcommand{\PERP}{{\scriptscriptstyle\perp}}
\newcommand{\PARA}{{\text{\large$\scriptstyle\shortparallel$}}}

\newcommand{\KET}[1]{|\,#1\,\rangle}
\newcommand{\BRA}[1]{\langle\,#1\,|}

\begin{document}

\title{Geometric Algebra in Quantum Information Processing}

\author{Timothy F.~Havel}
\address{T.~F.~Havel, BCMP, Harvard Medical School, Boston, MA, USA}
\curraddr{NED, Massachusetts Institute of Technology, Cambridge, MA, USA}
\email{tfhavel@mit.edu}
\thanks{\\ \hspace*{18pt}TH is supported by ARO grant DAAD19-01-1-0519,
DAAG55-97-1-0342 from DARPA/MTO, and DARPA/DSO grant MDA972-01-1-0003}

\author{Chris J.~L.~Doran}
\address{C.~Doran, MRAO, Cavendish Laboratory,
Madingley Road, Cambridge, CB3 0HE, United Kingdom}
\email{C.Doran@mrao.cam.ac.uk}
\thanks{CD is supported by an EPSRC Advanced Fellowship}

\date{\today}

\begin{abstract}
This paper develops a geometric model for
coupled two-state quantum systems (qubits)
using geometric (aka Clifford) algebra.
It begins by showing how Euclidean spinors
can be interpreted as entities in the
geometric algebra of a Euclidean vector space.
This algebra is then lifted to Minkowski
space-time and its associated geometric algebra,
and the insights this provides into how
density operators and entanglement behave
under Lorentz transformations are discussed.
The direct sum of multiple copies of space-time induces
a tensor product structure on the associated algebra,
in which a suitable quotient is isomorphic to the matrix
algebra conventionally used in multi-qubit quantum mechanics.
Finally, the utility of geometric algebra in
understanding both unitary and nonunitary quantum
operations is demonstrated on several examples
of interest in quantum information processing.
\end{abstract}

\maketitle

\section{Introduction}
Quantum mechanics attaches physical significance
to representations of the rotation group which differ
substantially from those studied in classical geometry.
Much of the mystery surrounding it is due to this fact.
The enormous interest recently generated by
proposals to build a \emph{quantum computer}
\cite{Lloyd:95,EkertJozsa:96,Steane:98a,WilliClear:99,Brooks:99,BenneDivin:00}
has focussed attention on the simplest possible quantum system:
a two-state system or \emph{qubit}.
Our understanding of qubits is based on two distinct
geometric models of their states and transformations:
\enlargethispage*{\baselineskip}
\begin{itemize}
\item A complex projective line
under the action of $\GRP{SU}(2)$.
\item A Euclidean unit $2$-sphere
under the action of $\GRP{SO}(3)$.
\end{itemize}
The first is used almost exclusively in fundamental
quantum physics, while the second (``classical'')
model is used extensively in certain applications,
e.g.~nuclear magnetic resonance (NMR) spectroscopy \cite{HaSoTsCo:00}.
In particular, in quantum computing a qubit represents a
binary $0$ or $1$ as its state corresponds to one of a pair of
conjugate (orthogonal) projective points $\KET{0}$ or $\KET{1}$.
These in turn correspond to a pair of diametrical points on
the unit sphere, which determine the alignment of the qubit
with or against the corresponding axis of quantization.

Formally, these two models are related by stereographic
projection of the Riemann (unit) sphere onto the Argand plane,
the points of which are the ratios of the homogeneous coordinates of
points on the projective line (see e.g.\ \cite{Altmann:86,FrescHiley:81}).
While this elegant construction describes the mapping
between the two representations in a geometric fashion,
it does not unite them in a single mathematical structure.
This paper provides an informal account of how
this is done by geometric (aka Clifford) algebra;
in addition, it describes an extension
of this formalism to multi-qubit systems,
and shows that it provides a concise and lucid
means of describing the operations of quantum
information processing \cite{SomCorHav:98,HaSoTsCo:00}.
Significantly, this extension is most naturally derived via
the geometric algebra of Minkowski space-time \cite{DorLasGul:93},
which has also been shown to be an efficient formalism within
which to study a very wide range of problems in classical
\cite{Hest$NF1:99,Jancewicz:89}, relativistic \cite{Hestenes:66,Baylis:96}
and fundamental quantum \cite{DoLaGuSoCh:96} physics.
More complete and rigorous accounts may be found in
these references, and in \cite{HaCoSoTs:00,Havel:01}.

\section{Euclidean Geometry and Spinors}
Let $\TEN R_3$ be a three-dimensional Euclidean
vector space whose inner product is denoted by
$(\VEC a, \VEC b) \mapsto \VEC a \cdot \VEC b$.
The Clifford or \emph{geometric algebra} of $\TEN R_3$
is the associative algebra generated by $\TEN R_3$ over
$\FLD R$ such that $\VEC a^2 = \| \VEC a \|^2 \equiv
\VEC a \cdot \VEC a$ for all $\VEC a \in \TEN R_3$.
This algebra will be referred to in the following as
the \emph{Pauli algebra}, and denoted by $\ALG G_3$.
The interesting thing about this algebra is its
geometric interpretation, which will now be described.

To begin, note that every nonzero vector $\VEC a
\in \TEN R_3$ has an inverse $\VEC a / \| \VEC a \|^2$.
In addition, a simple application of the law of
cosines shows that the inner product of $\VEC a$
with any other vector $\VEC b \in \TEN R_3$ is given
by the symmetric part of the geometric product:
\begin{equation} \begin{split}
\HALF (\VEC{ab} + \VEC{ba}) ~=~ &
\HALF( (\VEC a + \VEC b)^2 - \VEC a^2 - \VEC b^2) \\ =~ &
\HALF( \|\VEC a + \VEC b\|^2 - \|\VEC a\|^2 - \|\VEC b\|^2)
~=~ \VEC a \cdot \VEC b
\end{split} \end{equation}
The antisymmetric part, by way of contrast,
is called the \emph{outer product},
and denoted by $(\VEC a, \VEC b) \mapsto
\VEC a \wedge \VEC b \equiv (\VEC{ab} - \VEC{ba}) / 2$.
Since the outer product of two vectors $\VEC a \wedge \VEC b$ is
invariant under inversion in the origin, it cannot itself be a vector.
The space $\langle \VEC a\wedge\VEC b \,|\, \VEC a,\VEC b \in \TEN R_3 \rangle$
therefore carries an inequivalent representation of the orthogonal group
$\GRP O(3)$, and its elements are accordingly called \emph{bivectors}.
These are most naturally interpreted as oriented plane segments,
instead of oriented line segments like vectors in $\TEN R_{3\,}$.
If we similarly define the outer product of a vector
with a bivector and require it to be associative, i.e.
\begin{equation}
\VEC a \wedge (\VEC b \wedge \VEC c)
~\equiv~ \HALF (\VEC{abc} - \VEC{cba}) ~\equiv~
(\VEC a \wedge \VEC b) \wedge \VEC c
\end{equation}
($\VEC a, \VEC b, \VEC c \in \TEN R_3$),
then it can be shown via straightforward though
somewhat lengthy calculations that this product
of three vectors is totally antisymmetric, meaning
that the outer product generates the well-known exterior
algebra $\bigwedge \TEN R_3$ (cf.~\cite{Hest$NF1:99,Riesz:58}).
The outer product of three vectors is called a \emph{trivector},
and (since it changes sign under inversion) is most appropriately
interpreted as an oriented space segment or volume element.

The general properties of inner and outer products
in the geometric algebras of arbitrary metric vector
spaces can be worked out along these lines in
a coordinate-free fashion \cite{HesteSobcz:84}.
The remainder of this section will focus
on how the Pauli algebra is used to
describe the quantum mechanics of qubits.
In this application it is more common
to work with a fixed orthonormal basis
$\SIG_\LAB{x}, \SIG_\LAB{y}, \SIG_\LAB{z} \in \TEN R_3$.
Quantum mechanics, however, views these basis vectors
in a very different way from that taken above,
in that they are regarded as \emph{operators}
on a two-dimensional Hilbert space $\TEN H
\approx \FLD C^2$ (see e.g.~\cite{Sakurai:94}).
These operators, in turn, are usually
identified with the Pauli matrices
\begin{equation}
\SIG_\LAB{x} \;\leftrightarrow\; \UL\SIG_\LAB{\,x} \;\equiv\;
\begin{bmatrix} 0&1\\ 1&0 \end{bmatrix} ,~
\SIG_\LAB{y} \;\leftrightarrow\; \UL\SIG_\LAB{\,y} \;\equiv\;
\begin{bmatrix} 0&-\imath\\ \imath&0 \end{bmatrix} ,~
\SIG_\LAB{z} \;\leftrightarrow\; \UL\SIG_\LAB{\,z} \;\equiv\;
\begin{bmatrix} 1&0\\ 0&-1 \end{bmatrix} ,
\end{equation}
where $\imath$ is an imaginary unit ($\imath^2 = -1$),
the underline signifies that the associated symbol is a matrix,
and throughout this paper the symbol ``$\leftrightarrow$''
should be read as ``is represented by'' or ``is equivalent to''.
The connection between the two viewpoints lies the fact
that these matrices satisfy the defining relations of
the \emph{abstract} Pauli algebra $\ALG G_3$, namely
\begin{equation} \label{eq:pauli_def}
(\SIG_\mu)^2 ~=~ \EMB1 ~\leftrightarrow~ 1 ~,\quad
\SIG_\mu\SIG_\nu ~=~ - \SIG_\nu\SIG_\mu
\quad (\mu, \nu \in \{\LAB{x},\LAB{y},\LAB{z}\}, \mu \ne \nu) ~,
\end{equation}
and hence constitute a faithful matrix representation of it.
This shows, in particular, that $\ALG G_3$ is
$8$-dimensional as a \emph{real} linear space.%
\footnote{ In the quantum mechanics literature,
the notation $\UL{\VEC a} \cdot \UL{\vec{\SIG}}\,$
is often used for $\sum_\mu a_\mu \UL{\SIG}_\mu\,$.
Because $\VEC a$ is geometrically just a vector
in $\TEN R_3$ (\emph{not} a matrix for it in a
basis-dependent representation of $\ALG G_{3\,}$),
this is an abuse of the dot-notation for the
Euclidean inner product, which is otherwise perhaps
the most consistently used notation in all of science.
This abuse of notation will not be perpetrated in this paper.}

In most physical situations, these operators
(times $\hbar$) represent measurements of
the intrinsic angular momentum of the qubits,
and hence are regarded as generators of
rotations in the Lie algebra $\GRP{so}(3)$
over $\FLD C$ satisfying the commutator relation
\begin{equation}
\HALF [\,\SIG_\LAB{x}, \SIG_\LAB{y}\,] ~=~ \imath \SIG_\LAB{z} ~,
\end{equation}
and its cyclic permutations.
In terms of geometric algebra, the left-hand
side is just the outer product of the vectors.
The right-hand side is somewhat harder to interpret,
because the Pauli algebra is defined over the real numbers.
The trick is to observe that, in terms of the matrix representation,
$\UL\SIG_\LAB{\,x\,}\UL\SIG_\LAB{\,y\,}\UL\SIG_\LAB{\,z} = \imath \UL{\VEC1}$.
Thus by interpreting the abstract imaginary $\imath$ as the
trivector $\UPS \equiv \SIG_\LAB{x\,}\SIG_\LAB{y\,}\SIG_\LAB{z\,}
= \SIG_\LAB{x\,}\wedge\SIG_\LAB{y\,}\wedge\SIG_\LAB{z\,}$,
the angular momentum relations become a triviality:
\begin{equation}
\SIG_\LAB{x} \wedge \SIG_\LAB{y} ~=~ \SIG_\LAB{x\,} \SIG_\LAB{y} ~=~
\SIG_\LAB{x\,} \SIG_\LAB{y} {(\SIG_\LAB{z})}^2 ~=~ \UPS \SIG_\LAB{z}
\end{equation}
More generally, the vector cross product is related to the outer product by
\begin{equation}
\VEC a \times \VEC b ~=~ -{\textstyle\frac{\UPS}2}( \VEC{ab} - \VEC{ba} )
~=~ -\UPS(\VEC a \wedge \VEC b) ~,
\end{equation}
from which it may be seen that multiplication
by the unit trivector $\UPS$ maps vectors
to orthogonal bivectors and vice versa.
Since they span a one-dimensional space
but change sign under inversion in the origin,
trivectors can also be regarded as \emph{pseudo-scalars}.
Perhaps the most important thing which geometric algebra
contributes to physics are geometric interpretations for
the imaginary units which it otherwise uses blindly.

If we denote the induced bivector basis by
\begin{equation}
\VEC I ~\equiv~ \SIG_\LAB{y}\wedge\SIG_\LAB{z} ~,\quad
\VEC J ~\equiv~ \SIG_\LAB{z}\wedge\SIG_\LAB{x} ~,\quad
\VEC K ~\equiv~ \SIG_\LAB{x}\wedge\SIG_\LAB{y} ~,
\end{equation}
it is readily seen that these basis
bivectors likewise square to $-1$.
On multiplying the angular momentum generating
relations through by $-1 = \UPS^2$, we obtain
\begin{equation}
\VEC{JI} ~=~ \VEC K ~,\quad
\VEC{IK} ~=~ \VEC J ~,\quad
\VEC{KJ} ~=~ \VEC I ~,\quad
\text{and}~\VEC{KJI} ~=~ -1 ~.
\end{equation}
This shows that these basis bivectors generate a subalgebra of $\ALG G_3$
isomorphic to Hamilton's quaternions \cite{Altmann:86,Altmann:89},
which is also known as the \emph{even subalgebra} $\ALG G_3^+$
(since it is generated by the products of even numbers of vectors).
It is well-known that the quaternions' multiplicative group
is $\FLD R^* \oplus \GRP{SU}(2)$, which implies that the
even subalgebra should be closely related to rotations.
This relationship will now be worked out explicitly.

Consider the result of conjugating a
vector $\VEC x$ by a vector $\VEC a$, i.e.
\begin{equation}
\VEC{axa}^{-1} ~=~ \VEC a(\VEC x_\PARA +
\VEC x_\PERP)\VEC a^{-1} ~=~ \VEC{aa}^{-1}
\VEC x_\PARA - \VEC{aa}^{-1} \VEC x_\PERP ~,
\end{equation}
where we have split $\VEC x = (\VEC x \cdot \VEC a^{-1} + \VEC x
\wedge \VEC a^{-1}) \VEC a \equiv \VEC x_\PARA + \VEC x_\PERP$
into its parts parallel and perpendicular to $\VEC a$.
This shows that $-\VEC{axa}^{-1}$ is the \emph{reflection}
of $\VEC x$ in the plane orthogonal to $\VEC a$.
From the well-known fact that the composition of two
reflections is a rotation by \emph{twice} the lessor angle
between their planes and about these planes' line of intersection,
it follows that conjugating a vector by an element
of the even subalgebra just rotates it accordingly:
\begin{equation}
(\VEC{ba})\,\VEC{x}\,{(\VEC{ba})}^{-1} ~=~
\VEC{ba}\,\VEC{x}\,\VEC{a}^{-1}\VEC{b}^{-1} ~=~
\frac{\VEC{ba}\,\VEC{x}\,\VEC{ab}}{\|\VEC a\|^2 \| \VEC b \|^2}
\end{equation}
Let $\VEC u \equiv \VEC a / \| \VEC a \|$,
$\VEC v \equiv \VEC b / \| \VEC b \|$ and
$\VEC R \equiv \VEC{vu}$ be the corresponding \emph{unit} quaternion.
Then $\VEC R = \cos(\theta/2) - \UPS \VEC r \sin(\theta/2)$
where $\cos(\theta/2) = \VEC u \cdot \VEC v$ and $\UPS \VEC r
\equiv \VEC u \wedge \VEC v / \| \VEC u \wedge \VEC v \|$.
Moreover, the inverse ${(\VEC{vu})}^{-1}$ is now simply
the \emph{reverse} $\VEC{uv} \equiv {(\VEC{vu})}^\dag$,
which in turn corresponds to the \emph{conjugate} quaternion
${\VEC R}^\dag \equiv \cos(\theta/2) + \UPS \VEC r \sin(\theta/2)$.
This reversal operation on $\ALG G_3^+$ extends
to a well-defined anti-automorphism of $\ALG G_3\,$,
which corresponds to Hermitian conjugation
in its representation by Pauli matrices.
On splitting $\VEC x$ into its parts parallel
$\VEC x_\PARA$ and perpendicular $\VEC x_\PERP$
to $\VEC r$ as above, the rotation may now be
written as $\VEC{R} \,\VEC x\, {\VEC R}^\dag = \VEC R
(\VEC x_\PARA + \VEC x_\PERP) {\VEC R}^\dag =
\VEC x_\PARA + \VEC x_\PERP {({\VEC R}^\dag)}\DAB{1.6ex}^2$
\begin{equation} \begin{split}
=~ & \VEC x_\PARA + \VEC x_\PERP
(\cos^2(\theta/2) - \sin^2(\theta/2) +
2 \UPS \VEC r \cos(\theta/2) \sin(\theta/2)) \\
=~ & \VEC x_\PARA + \VEC x_\PERP
( \cos(\theta) + \UPS \VEC r \sin(\theta) ) ~,
\end{split} \end{equation}
and so may be viewed as multiplication of
$\VEC x_\PERP$ by the ``complex number''
$\cos(\theta) + \UPS \VEC r \sin(\theta)$ in the
Argand plane defined by the bivector $\UPS \VEC r$.

By collecting even and odd powers in its Taylor series,
it may be seen that any unit quaternion can be written as the
exponential of a bivector orthogonal to the axis of rotation $\VEC r$:
\begin{equation}
e^{-\UPS\VEC r\theta/2} ~=~
1 - \UPS\VEC r \, {\textstyle\frac{\theta}2} -
\HALF \left({\textstyle\frac{\theta}2}\right)^2 +
\cdots ~=~ \cos(\theta/2) - \UPS\VEC r\sin(\theta/2)
\end{equation}
This is formally analogous to a complex exponential,
and is also in accord with our previous observation that
the space of bivectors is isomorphic to the Lie algebra
$\GRP{su}(2) \approx \GRP{so}(3)$ under the commutator product.
The pair $[\cos(\theta/2); \sin(\theta/2)\VEC r]$
are known as \emph{Euler-Rodrigues} parameters
for the rotation group $\GRP{SO}(3)$; since
$[-\cos(\theta/2); -\sin(\theta/2)\VEC r]$ determines
the same rotation, this parametrization is two-to-one.
A one-to-one parametrization is obtained
from the \emph{outer exponential}, i.e.
\begin{equation}
\wedge^{-\UPS \VEC r \tau} ~=~ 1 - \UPS \VEC r \tau - \HALF
\VEC r \wedge \VEC r \tau^2 + \cdots ~=~ 1 - \UPS \VEC r \tau \quad
\text{(since $\VEC r \wedge \VEC r = 0$)} ~.
\end{equation}
The squared norm of this outer exponential is $1 + \tau^2$,
so that the normalized outer exponential equals the
usual exponential if we set $\tau = \tan(\theta/2)$.
Because $\VEC t \equiv \tan(\theta/2) \VEC r$
is the four-dimensional stereographic projection
of $[\cos(\theta); \sin(\theta)\VEC r]$
from $\theta = \pi$, it has been called the
\emph{stereographic} parameter for $\GRP{SO}(3)$.
Note, however, that this parametrization
does not include rotations by $\pi$.

Another two-to-one parametrization of rotations is given by the
\emph{Cayley-Klein} parameters $[\psi_1; \psi_2] \in \FLD C^2$, where
\begin{equation} \begin{split}
\psi_1 ~\equiv~ & \cos(\theta/2) -
\imath\sin(\theta/2)(\VEC r \cdot \SIG_\LAB{z}) ~, \\
\psi_2 ~\equiv~ & \sin(\theta/2)(\VEC r \cdot \SIG_\LAB{x})
+ \imath\sin(\theta/2)(\VEC r \cdot \SIG_\LAB{y}) ~.
\end{split} \end{equation}
The corresponding $\GRP{SU}(2)$ matrix is simply
\begin{equation}
\UL{\EMB\Psi} ~=~ \begin{bmatrix} \psi_1 & -\psi_2^* \\
\psi_2 & \psi_1^* \end{bmatrix} ~.
\end{equation}
It follows that the complex column vector
$\KET{\psi} \equiv [\psi_1; \psi_2]$ itself transforms
under left-multiplication with matrices in $\GRP{SU}(2)$,
which is commonly described in quantum
mechanics by calling it a \emph{spinor}.
In particular, the spinors $\KET{0} \equiv [ 1; 0 ]$
and $\KET{1} \equiv [ 0; 1 ]$ are those commonly
used in quantum computing to store binary information.
Since the Cayley-Klein parameters uniquely determine
the $\GRP{SU}(2)$ matrix, however, we can just as
well regard spinors as entities \emph{in} $\GRP{SU}(2)$,
e.g.~$\KET{0} \leftrightarrow \UL{\VEC 1}$ and
$\KET{1} \leftrightarrow -\imath\UL{\SIG}_\LAB{\,y}$.
The usual action of $\GRP{SU}(2)$ on spinors then becomes
the left-regular action of $\GRP{SU}(2)$ on itself.

The representation of $\GRP{SU}(2)$ used above
depends upon the choice of coordinate system:
Changing to a different the coordinate system gives
a different (though equivalent) representation.
Recalling that $\GRP{SU}(2)$ is isomorphic
to the multiplicative group of unit elements
(quaternions) in the even subalgebra $\ALG{G}_3^+$,
a coordinate-free or \emph{geometric} interpretation of spinors
is obtained by regarding them as elements of $\ALG{G}_3^+$ itself.
This interpretation of spinors as entities in
ordinary Euclidean geometry was first pointed out
by Hestenes over thirty years ago \cite{Hestenes:66},
but physicists persist in putting operators
and operands into separate spaces, and in
working with a matrix representation instead
of directly with the geometric entities themselves.
The perceived nonintuitive nature of quantum mechanics
is due in large part to the resulting confusion over
the geometric meaning of the objects with which it deals,
which is spelled out explicitly in geometric algebra.

As another example, consider how the \emph{density operator}
of an ``ensemble'' of qubits can be interpreted in geometric algebra.
This operator $\RHO$ is usually defined via a matrix
representation as $\UL{\RHO} \equiv \OL{\KET{\psi}\BRA{\psi}}$,
where the overline denotes the average over the ensemble.
As first observed by von Neumann, this matrix contains
all the information needed to compute the ensemble average
expectation values of the qubit observables, since
\begin{equation} \label{eq:rho1_obs}
\OL{\BRA{\psi}\UL{\SIG}_{\,\mu}\KET{\psi}} ~=~
\OL{\TR(\UL{\SIG}_{\,\mu}\KET{\psi}\BRA{\psi})} ~=~
\TR(\UL{\SIG}_{\,\mu}\OL{\KET{\psi}\BRA{\psi}})
~=~ \TR(\UL{\SIG}_{\,\mu} \UL{\RHO} )
\end{equation}
($\mu \in \{\LAB{x},\LAB{y},\LAB{z}\}$).
To translate this into geometric algebra, we set the second column of
$\UL{\EMB\Psi}$ to zero by right-multiplying it by the idempotent matrix
$\UL{\VEC E}_{\,+} ~\equiv~ (\UL{\VEC 1} + \UL\SIG_\LAB{\,z})/2$, i.e.
\begin{equation}
\UL{\EMB\Psi} \, \UL{\VEC E}_{\,+} ~\equiv~ 
\begin{bmatrix} \psi_1 & -\psi_2^* \\
\psi_2 & \psi_1^* \end{bmatrix}
\begin{bmatrix} 1&0 \\ 0&0 \end{bmatrix}
~=~ \begin{bmatrix} \psi_1 & 0 \\
\psi_2 & 0 \end{bmatrix} ~.
\end{equation}
This corresponds to projecting $\EMB\Psi \in
\ALG G_3^+$ onto a \emph{left-ideal} in $\ALG G_3\,$,
and allows the dyadic product $\KET{\psi}\BRA{\psi}$
in Eq.~(\ref{eq:rho1_obs}) to be written as:
\begin{equation}
\KET{\psi}\BRA{\psi} ~=~
\begin{bmatrix} \psi_1 & 0 \\
\psi_2 & 0 \end{bmatrix}
\begin{bmatrix} \psi_1^* & \psi_2^* \\
0 & 0 \end{bmatrix} ~\equiv~
(\UL{\EMB\Psi} \, \UL{\VEC E}_{\,+})
{(\UL{\EMB\Psi} \, \UL{\VEC E}_{\,+})}^{\dag}
\end{equation}
Thus the interpretation of the density
operator in geometric algebra is
\begin{equation} \label{eq:den_op_def}
\RHO ~=~ \OL{(\EMB\Psi\VEC E_+){(\EMB\Psi\VEC E_+)}^\dag}
~=~ \OL{\EMB\Psi\VEC E_+{\EMB\Psi}^\dag} ~=~ \HALF\left(
1 + \OL{\EMB\Psi\SIG_\LAB{z}{\EMB\Psi}^\dag}\right)
\end{equation}
(cf.~\cite{SomLasDor:99}).
The vector part $\VEC p \equiv {\langle\RHO\rangle}_1
= \OL{ \EMB\Psi\SIG_\LAB{z}{\EMB\Psi}^\dag}$
is called the \emph{polarization vector}
(in optics, its components are known as the
\emph{Stokes parameters} \cite{BaBoDeHu:93},
while in NMR it is known as the \emph{Bloch vector} after
the pioneer of NMR who rediscovered it \cite{Bloch:46}).
Its length is $\| \VEC p \| \le 1$ with
equality if and only if all members of
the ensemble are in the same state $\EMB\Psi$.
In this case the ensemble is said to be in a \emph{pure} state,
and the density operator is itself an idempotent $(1 + \VEC p)/2$,
where $\VEC p \equiv \EMB\Psi\SIG_\LAB{z}{\EMB\Psi}^\dag$.
For an ensemble in a general \emph{mixed} state,
the length of the ensemble-average polarization vector
measures the degree of alignment among the (unit length)
polarization vectors of the individual members of the ensemble,
and is called the \emph{polarization} of the ensemble.

In many physical situations there is a natural reference direction;
for example, in NMR computing the qubits are spin $1/2$ atomic
nuclei whose intrinsic magnetic dipoles have been polarized by
the application of a strong magnetic field \cite{HaCoSoTs:00}.
From a geometric perspective, however, the density operator is just
the sum of a scalar and a vector, which for a pure state is related
to the corresponding ``spinor'' by rotation of a fixed reference vector
(conventionally taken to be $\SIG_\LAB{z}$ as above) by $\EMB\Psi$.
Since the trace in the standard matrix representation is
simply twice the scalar part ${\langle\,\UL{~}\,\rangle}_0$
of the corresponding expression in geometric algebra,
the ensemble-average expectation value
\begin{equation}
\HALF\, \TR( \UL{\SIG}_{\,\mu\,} \UL{\RHO} ) ~\leftrightarrow~
{\langle \SIG_{\mu\,} \RHO \rangle}_0 ~=~ {\langle \SIG_\mu
\EMB\Psi \SIG_\LAB{z} {\EMB\Psi}^\dag \rangle}_0 ~=~
\SIG_\mu \cdot ( \EMB\Psi \SIG_\LAB{z} {\EMB\Psi}^\dag )
~=~ \SIG_\mu \cdot \VEC p
\end{equation}
is just the component of the polarization vector along the $\mu$-th axis.
Unlike the \emph{strong} measurements usually considered in quantum texts,
where measurement of $\SIG_\mu$ yields one of the random outcomes $\pm1$
with probabilities $(1 \pm \SIG_\mu \cdot \VEC p)/2$ and leaves
the system in the corresponding state $\VEC p = \pm \SIG_\mu\,$,
\emph{weak} measurements of ensemble-average expectation
values can be made with only negligible perturbations
to the ensemble as a whole \cite{Peres:93}.
This is in fact how quantum mechanical systems
are usually manifest at the macroscopic level!

To see how all this relates to conventional wisdom,
observe that the polarization vector of a pure state may be
written in terms of the Cayley-Klein parameters as
\begin{equation}
\VEC p ~=~ 2\Re(\psi_1^*\psi_2)\SIG_\LAB{x}
+ 2\Im(\psi_1^*\psi_2)\SIG_\LAB{y} +
(|\psi_1|^2 - |\psi_2|^2) \SIG_\LAB{z} ~.
\end{equation}
Its stereographic projection from $-\SIG_\LAB{z}$ onto
the $\SIG_\LAB{x}\SIG_\LAB{y}$ plane is therefore
\begin{equation}
\frac{ 2\Re(\psi_1^*\psi_2)\SIG_\LAB{x} +
2\Im(\psi_1^*\psi_2)\SIG_\LAB{y} }
{ 1 + |\psi_1|^2 - |\psi_2|^2 } ~.
\end{equation}
Multiplying by $\SIG_\LAB{x}$ and simplifying the
denominator using $|\psi_1|^2 + |\psi_2|^2 = 1$ yields
\begin{equation}
\frac{ \Re(\psi_1^*\psi_2) + \Im(\psi_1^*\psi_2) \VEC K }{ |\psi_1|^2 }
\end{equation}
where $\VEC K = \SIG_\LAB{x}\SIG_\LAB{y}$
is a square-root of $-1$.
This is the same as the ratio $\psi_2/\psi_1$ save for
the use of $\imath$ instead of $\VEC K$ as the imaginary unit,
which explains formally why $\GRP{SO}(3)$ acts on the polarization
vector in the same way that $\GRP{SU}(2)$ acts on the ratio of
the Cayley-Klein parameters \cite{Altmann:86,FrescHiley:81}.

\section{Space-Time Geometry and Multiparticle Spinors}
The above interpretations apply only to single qubits
(or to ensembles consisting of noninteracting and identical qubits).
Extending them to systems of interacting and distinguishable
qubits may be done in a physically significant fashion by
considering the geometric algebra of \emph{space-time}
(or Minkowski space) $\TEN R_{1,3\,}$.
This algebra, known as the \emph{Dirac} algebra
and denoted by $\ALG G_{1,3}$, may be defined by
the generating relations among an orthonormal
basis analogous to Eq.~(\ref{eq:pauli_def}):
\begin{equation} \begin{split}
& \GAM_\LAB{t}^2 ~=~ 1 ~,\quad \GAM_\mu^2 ~=~ -1
\quad( \mu \in \{ \LAB{x}, \LAB{y}, \LAB{z} \}) ~, \\
& \GAM_\mu \GAM_\nu ~=~ -\GAM_\nu \GAM_\mu \quad( \mu, \nu
\in \{ \LAB{t}, \LAB{x}, \LAB{y}, \LAB{z} \}, \mu\ne\nu )
\end{split} \end{equation}
The corresponding geometric algebra separates
into five inequivalent representations under the
action of the full Lorentz group $\GRP O(1,3)$, i.e.
\begin{equation} \begin{split}
\langle 1 \rangle \quad &
	(\text{scalars, $1$-dimensional}) \\
\langle \GAM_\mu \rangle \quad &
	(\text{vectors, $4$-dimensional}) \\
\langle \GAM_\mu\GAM_\nu \rangle \quad &
	(\text{bivectors, $6$-dimensional}) \\
\langle \GAM_\mu\GAM_\nu\GAM_\eta \rangle \quad &
	(\text{trivectors, $4$-dimensional}) \\
\langle \GAM_\LAB{t}\GAM_\LAB{x}\GAM_\LAB{y}\GAM_\LAB{z} \rangle \quad &
	(\text{pseudo-scalars, $1$-dimensional}) ~,
\end{split} \end{equation}
where $\mu,\nu,\eta \in \{ \LAB{t}, \LAB{x}, \LAB{y}, \LAB{z} \}$
with $\mu \ne \nu \ne \eta \ne \mu$, for a total dimension of $16$.

The important point for our purposes is that the even
subalgebra of the Dirac algebra $\ALG G_{1,3}^+$ is
isomorphic to the Pauli algebra $\ALG G_3$ \cite{Hestenes:66}.
This isomorphism may be constructed by choosing bases
$\GAM_\mu \in \ALG G_{1,3}$ and $\SIG_\mu \in \ALG G_3$,
and defining an invertible linear mapping by
\begin{equation}
\SIG_\mu \in \ALG G_3 ~\leftrightarrow~ \GAM_\mu \GAM_\LAB{t} \in
\ALG G_{1,3}^+ \quad( \mu \in \{ \LAB{x}, \LAB{y}, \LAB{z} \}) ~.
\end{equation}
These so-called \emph{relative spatial vectors} $\GAM_\mu\GAM_\LAB{t}$
satisfy the relations in Eq.~(\ref{eq:pauli_def}), since
\begin{equation} \begin{split}
(\SIG_\mu)^2 ~\leftrightarrow~ &
(\GAM_\mu \GAM_\LAB{t})^2 ~=~ -\GAM_\mu
(\GAM_\LAB{t})^2 \GAM_\mu ~=~ -(\GAM_\mu)^2 ~=~ 1 \\
\SIG_\mu \SIG_\nu ~\leftrightarrow~ &
(\GAM_\mu\GAM_\LAB{t})(\GAM_\nu\GAM_\LAB{t})
~=~ \GAM_\nu (\GAM_\mu \GAM_\LAB{t}) \GAM_\LAB{t}
~=~ -(\GAM_\nu\GAM_\LAB{t})(\GAM_\mu\GAM_\LAB{t}) \\
\leftrightarrow~ & -\SIG_\nu \SIG_\mu \quad
( \mu, \nu \in \{ \LAB{x}, \LAB{y}, \LAB{z} \},
~\mu\ne\nu ) ~,
\end{split} \end{equation}
and hence generate an algebra isomorphic to $\ALG G_3$.
As bivectors in $\ALG G_{1,3}$, however,
they also generate $\ALG G_{1,3}^+$, since
\begin{equation}
\GAM_\mu \GAM_\nu ~=~ \GAM_\mu (\GAM_\LAB{t})^2
\GAM_\nu ~=~ -(\GAM_\mu\GAM_\LAB{t})(\GAM_\nu\GAM_\LAB{t})
~\leftrightarrow~ -\SIG_\mu \SIG_\nu
\end{equation}
($\mu, \nu \in \{ \LAB{x}, \LAB{y}, \LAB{z} \}$, $\mu\ne\nu$),
and similarly
\begin{equation}
\UPS ~\equiv~ \GAM_\LAB{t}\GAM_\LAB{x}\GAM_\LAB{y}\GAM_\LAB{z} ~=~
(\GAM_\LAB{x}\GAM_\LAB{t})(\GAM_\LAB{y}\GAM_\LAB{t})(\GAM_\LAB{z}\GAM_\LAB{t})
~\leftrightarrow~ \SIG_\LAB{x}\SIG_\LAB{y}\SIG_\LAB{z} ~.
\end{equation}
Thus $\GAM_\mu\GAM_\LAB{t} \leftrightarrow \SIG_\mu$
($\mu \in \{ \LAB{x}, \LAB{y}, \LAB{z} \}$)
induces an algebra isomorphism as claimed,
and when the bases are understood we may
identify $\SIG_\mu \equiv \GAM_\mu\GAM_\LAB{t}$.

The choice of time-like vector $\GAM_\LAB{t} \in \TEN R_{1,3}$
in fact determines an inertial frame up to spatial rotation,
in which the \emph{time} $t$ and \emph{place} $\VEC s$
of an event $\MAT e$ in that frame are given by
\begin{equation}
t \,+\, \VEC s ~=~ \MAT e \cdot \GAM_\LAB{t} \,+\,
\MAT e \wedge \GAM_\LAB{t} ~=~ \MAT e \, \GAM_\LAB{t}
\end{equation}
(note that upright case is used for the
space-time vector $\MAT e \in \TEN R_{1,3}$).
Thus the invariant interval between events
separated by the space-time vector $\MAT e$ is
$\MAT e^2 = \MAT e\, \GAM_\LAB{t}^2\, \MAT e =
(t + \VEC s) (t - \VEC s) = t^2 - \VEC s^2$ as usual,
while the relative velocity between events
whose space-time velocities are $\GAM_\LAB{t}$
and $\MAT v \equiv \partial \MAT e / \partial \tau$ is
\begin{equation}
\VEC v ~=~
\frac{\partial \VEC s}{\partial t} ~=~
\frac{\partial \VEC s}{\partial \tau} \,
\, \frac{\partial \tau}{\partial t} ~\equiv~
\left( \frac{\partial \MAT e}{\partial \tau} \wedge
\GAM_\LAB{t} \right) \left( \frac{\partial \MAT e}
{\partial \tau} \cdot \GAM_\LAB{t} \right)^{-1} ~=~
\frac{\MAT v \wedge \GAM_\LAB{t}}
{\MAT v \cdot \GAM_\LAB{t}} ~,
\end{equation}
so that $\VEC v\cdot\GAM_\LAB{t}$ lies
on an affine hyperplane in space-time.

A great deal of physics can be done in a manifestly
Lorentz covariant fashion using the Dirac algebra.
For example, the electromagnetic field at a given
point in space-time corresponds to an arbitrary
bivector $\MAT F \in \bigwedge_2 \TEN R_{1,3}$,
called the \emph{Faraday bivector}, and the
covariant form of the Lorentz force equation is
\begin{equation}
m\, \dot{\MAT v} ~=~ q\, \MAT F \cdot \MAT v ~,
\end{equation}
where $m$ is the rest mass, $q$ the
charge and $\MAT v$ the space-time velocity.
(This is another example of the general rule
that, in geometric algebra, the generators
of motion are bivectors \cite{DoHeSoVA:93}.)
The usual frame-dependent form is recovered by
splitting the quantities in this equation by
$\GAM_\LAB{t}$ as above \cite{Jancewicz:89};
in particular, the Faraday bivector splits
into an electric and a magnetic field as
$\MAT F \equiv \VEC E + \UPS\VEC B$, where
\begin{equation}
\VEC E ~=~ (\MAT F \cdot \GAM_\LAB{t}) \GAM_\LAB{t}
\quad\text{and}\quad
\UPS\VEC B ~=~ (\MAT F \wedge \GAM_\LAB{t}) \GAM_\LAB{t} ~.
\end{equation}
The space-time reverse will be denoted by a tilde,
e.g.~in the present case $\tilde{\MAT F} = -\MAT F$.
This is related to the spatial (or Pauli) reverse by $\MAT F^\dag
= \VEC E - \UPS\VEC B = \GAM_\LAB{t\,} \tilde{\MAT F}_{\,} \GAM_\LAB{t}$.
Both operations agree on the Pauli-even subalgebra, but the spatial reverse
\emph{not} Lorentz coveriant since it depends on a particular $\GAM_\LAB{t\,}$.

Returning to our previous discussion of the density operator,
we observe that the space-time form of the density operator
of a single qubit polarized along $\LAB{z}$ can be written as
\begin{equation}
\RHO ~=~ \HALF ( 1 + \alpha \SIG_\LAB{z} ) ~=~ \HALF
( \GAM_\LAB{t} + \alpha \GAM_\LAB{z} ) \GAM_\LAB{t}
~\equiv~ \EMB\varrho \GAM_\LAB{t} ~,
\end{equation}
where $-1 \le \alpha \le 1$ is the polarization and
$\GAM_\LAB{t}$ determines the local inertial frame.
It follows that the Lorentz covariant form of the density
operator is a time-like vector $\EMB\varrho \in \TEN R_{1,3}$.
Under a Lorentz boost $\VEC L = \exp(-\lambda\SIG_\LAB{z}/2)
\in \GRP{SO}(1,3)$ along $\SIG_\LAB{z\,}$, therefore,
the relativistic density operator $\EMB\varrho$
transforms to $\EMB\varrho' ~\equiv~ \HALF \VEC L
(\GAM_\LAB{t} + \alpha \GAM_\LAB{z}) \tilde{\VEC L} ~=$
\begin{equation}
\HALF\left( \cosh(\lambda) \GAM_\LAB{t} - \sinh(\lambda) \GAM_\LAB{z} +
\alpha( \cosh(\lambda) \GAM_\LAB{z} - \sinh(\lambda) \GAM_\LAB{t} ) \right) ~.
\end{equation}
This implies that in the unaccelerated frame
(with renormalization by $\EMB\varrho' \cdot \GAM_\LAB{t}$),
\begin{equation} \begin{split}
\RHO' ~=~ & \frac{\EMB\varrho' \, \GAM_\LAB{t}}
{\EMB\varrho' \cdot \GAM_\LAB{t}} ~=~
\frac{\EMB\varrho' \cdot \GAM_\LAB{t} + \EMB\varrho' \wedge \GAM_\LAB{t}}
{\EMB\varrho' \cdot \GAM_\LAB{t}} \\
=~ & \frac12 \left( 1 \,+\, \frac{ \alpha \cosh(\lambda) - \sinh(\lambda)
}{ \cosh(\lambda) - \alpha \sinh(\lambda) } \, \SIG_\LAB{z} \right) ~.
\end{split} \end{equation}
It follows that the polarization itself transforms as
\begin{equation}
\alpha' ~=~ \frac{\alpha \cosh(\lambda) - \sinh(\lambda)}
{\cosh(\lambda) - \alpha \sinh(\lambda)} ~.
\end{equation}
If we assume the qubit is at equilibrium with a heat bath,
statistical mechanics tells us that $\alpha = \tanh(-\beta\epsilon/2)$
where $\beta = 1/(k_\LAB{B}T)$ is the inverse temperature
and $\epsilon \in \FLD R$ is the energy difference between
the $\KET{0}$ and $\KET{1}$ states \cite{Tolman:38}.
Then the addition formulae for $\cosh$ and $\sinh$ give
\begin{equation}
\alpha' ~=~ \tanh( -\beta\epsilon/2 - \lambda ) ~,
\end{equation}
so the apparent equilibrium polarization depends on velocity.
These results are not to be found in the classic treatise
on relativistic thermodynamics \cite{Tolman:34}.

We will now construct a Lorentz covariant \emph{multiparticle
theory} of qubit systems in the simplest possible way,
by taking a direct sum of copies of space-time
(regarded as a vector space, rather than an algebra),
one for each of the $N$ qubits, i.e.
\begin{equation}
{\textstyle\bigoplus}_{q=1}^N \, \left\langle \GAM_\LAB{t}^q,
\GAM_\LAB{x}^q,\GAM_\LAB{y}^q,\GAM_\LAB{z}^q \right\rangle ~,
\end{equation}
and considering the associated geometric algebra $\ALG G_{N,3N}$.
Then the even subalgebras of different particle spaces $p \ne q$
\emph{commute}, since (in any given bases)
\begin{equation}
\SIG_\mu^{\,p} \SIG_\nu^{\,q} ~=~
\GAM_\mu^p (\GAM_\nu^q \GAM_\LAB{t}^q ) \GAM_\LAB{t}^p
~=~ \SIG_\nu^{\,q} \SIG_\mu^{\,p}
\end{equation}
for all $\mu, \nu \in \{ \LAB{x}, \LAB{y}, \LAB{z} \}$,
so that the algebra generated by the even subalgebras is
isomorphic to a tensor product of these algebras, written as
\begin{equation}
(\ALG G_{1,3}^+)^{\otimes N} ~\approx~
\ALG G_3^{\otimes N} ~\equiv~ {(\ALG G_3)}^{\otimes N} ~.
\end{equation}
This construction of the tensor product was first
used by Clifford as a means of studying the tensor
products of quaternion algebras \cite{Clifford:1878};
van der Waerden has in fact called it a Clifford
algebra of the second kind \cite{vanderWaerden:85}.
As a means of justifying the tensor product
of nonrelativistic quantum mechanics in terms
of the underlying geometry of space-time, however,
it is a much more recent development \cite{DorLasGul:93}.

A key feature of quantum mechanics, which is needed
for quantum computers to be able to solve problems
more efficiently than their classical counterparts,
is an exponential growth in the dimension of the Hilbert space
of a multi-qubit system with the number of particles involved.
The complex dimension of the Hilbert space
$(\TEN H)^{\otimes N}$ of an $N$-qubit system is in fact $2^N$,
and the space of operators (linear transformations) on
$(\TEN H)^{\otimes N}$ therefore has \emph{real} dimension $2^{2N+1}$.
The above construction yields a space of ``operators''
$\ALG G_3^{\otimes N}$ whose real dimension
also grows exponentially, but as $2^{3N}$.
The extra degrees of freedom are due to the presence of a
different unit pseudo-scalar $\UPS^q$ in every particle space.
They can easily be removed by multiplying through by
an idempotent element called the \emph{correlator}:
\begin{equation}
\VEC C ~\equiv~ \HALF(1 - \UPS^1\UPS^2) \, \HALF(1 - \UPS^1\UPS^3)
~\cdots~ \HALF(1 - \UPS^1\UPS^N)
\end{equation}
This commutes with everything in $\ALG G_3^{\otimes N}$ and
satisfies $\UPS^p\UPS^q\VEC C = -\VEC C$ for $1 \le p,q \le N$,
so that multiplication by it homomorphically
maps $\ALG G_3^{\otimes N}\!$ onto an ideal
$\ALG G_3^{\otimes N} \!/ \VEC C$ wherein
all the unit pseudo-scalars have been identified,%
\footnote{The notation $\ALG G_3^{\otimes N} \!/ \VEC C$ 
is justified by the fact that the two-sided principle
ideal $\ALG G_3^{\otimes N}(\VEC C)$ generated by
$\VEC C$ is isomorphic to the quotient algebra 
$\ALG G_3^{\otimes N} \!/\, \FUN{ker}(\VEC C)$,
where $\FUN{ker}(\VEC C) \equiv \{ \VEC g \in
\ALG G_3^{\otimes N} \!\mid \VEC g \VEC C = 0 \}$.}
and which therefore has the correct dimension over $\FLD R$.
As a subalgebra, this ideal is in fact isomorphic
to the algebra of $2^N \times 2^N$ complex matrices,
and hence capable of describing all the states and
transformations of (ensembles of) $N$ qubit systems.
In the following, we shall generally omit $\VEC C$
from our expressions altogether, and use a single unit
imaginary $\UPS$ as in conventional quantum mechanics.

On the ``even'' subalgebra ${(\ALG G_3^+)}^{\otimes N}$,
multiplication by the correlator turns out to be an
algebra automorphism; this algebra can thus be written as
\begin{equation}
{(\ALG G_3^+)}^{\otimes N} ~\approx~
{(\ALG G_3^+)}^{\otimes N} \!/ \VEC C ~\approx~
{(\ALG G_3^{\otimes N} \!/ \VEC C)}^+ ~\approx~
{\ALG{SU}(2)}^{\otimes N} ~,
\end{equation}
where the ``$+$'' refers throughout to the subalgebra generated
by expressions which are invariant under inversion in the origin,
and ${\ALG{SU}(2)}\DAB{1.6ex}^{\otimes\smash N}$ to
the algebra generated over $\FLD R$ by the Kronecker
products of matrices in the group $\GRP{SU}(2)$.
This subalgebra has real dimension $2^{2N}$,
but is mapped onto a left-ideal of dimension $2^{N+1}$
by right-multiplication with another idempotent which is
given by the tensor product of those considered earlier, namely
\begin{equation}
\VEC E_{+} ~\equiv~ \VEC E_+^1 \VEC E_+^2 \cdots \VEC E_+^N ~,
\end{equation}
where $\VEC E_{\pm}^{\,q} ~\equiv~ (1 \pm \SIG_\LAB{z}^{\,q})/2$
for $q = 1,\ldots,N$.
Henceforth, the term ``even subalgebra'' will
refer to ${(\ALG G_3^+)}\DAB{1.6ex}^{\otimes\smash N}$
(suitably correlated) unless otherwise stated.

In terms of the usual matrix representation,
right-multiplication of an element of the even
subalgebra $\UL{\EMB\Psi}$ by $\UL{\VEC E}_{\,+}$
likewise sets all but the first column to zero,
so that $\EMB\Psi\VEC E_{+}$ transforms like
a ``spinor'' in $\TEN H^{\otimes N}$ under
left-multiplication by single particle rotations
${\VEC R}^q \in {(\ALG G_3^+)}\DAB{1.6ex}^{\otimes\smash N}$.
Unlike the single particle case, however, this one column
does not uniquely determine an element of the even subalgebra
${(\ALG G_3^+)}\DAB{1.6ex}^{\otimes\smash N}\!/\VEC C$.
What has been proposed instead \cite{DorLasGul:93} is to
use the fact that $\VEC E_{+}$ ``absorbs'' $\SIG_\LAB{z}$'s
to distribute copies of the latter across the correlator,
converting it to what will here be called the
\emph{directional} correlator $\VEC D$, i.e.
\begin{equation}
\EMB\Psi \VEC C \VEC E_{+} ~=~ \EMB\Psi \VEC C
\left({(\SIG_\LAB{z}^{\,1})}^{N-1} \SIG_\LAB{z}^{\,2}
\cdots \SIG_\LAB{z}^{\,N}\right) \VEC E_{+}
~=~ \EMB\Psi \VEC D \VEC E_{+} ~,
\end{equation}
where
\begin{equation}
\VEC D ~\equiv~
\HALF(1 - \UPS^1\SIG_\LAB{z}^{\,1}\UPS^2\SIG_\LAB{z}^{\,2}) \,
\HALF(1 - \UPS^1\SIG_\LAB{z}^{\,1}\UPS^3\SIG_\LAB{z}^{\,3}) ~\cdots~
\HALF(1 - \UPS^1\SIG_\LAB{z}^{\,1}\UPS^N\!\SIG_\LAB{z}^{\,N}) ~.
\end{equation}
It can be shown that right-multiplication by $\VEC D$,
unlike $\VEC C$, reduces the dimensionality to $2^{N+1}$,
thereby permitting the objects in this \emph{reduced} even
subalgebra ${(\ALG G_3^+)}\DAB{1.6ex}^{\otimes\smash N} \!/ \VEC D$
to be regarded as spinors,
analogous to $\ALG G_3^+$ for a single qubit.
In the corresponding left-ideal, $\VEC K \equiv
\UPS^1\SIG_\LAB{z}^{\,1} \VEC D \leftrightarrow \cdots
\leftrightarrow \UPS^N\!\SIG_\LAB{z}^{\,N} \VEC D$
serves as the unit imaginary, since $\VEC K^2 = -\VEC D$,
but is required to always operate from the \emph{right}.
Henceforth, unless otherwise mentioned, we will regard
spinors ${(\ALG G_3^+)}\DAB{1.6ex}^{\otimes\smash N} \!/ \VEC D
= ({(\ALG G_3^+)}\DAB{1.6ex}^{\otimes\smash N} \!/ \VEC C) / (\VEC{CD})$
as a left-ideal in the $\VEC C$-correlated even subalgebra,
drop both $\VEC C$ and the superscripts on the $\UPS$'s as above,
and use $\VEC D$ as a short-hand for $\VEC{CD} = \VEC{DC}$.

In the case of two qubits, for example,
the identifications are induced by $\VEC D$ are
\begin{alignat}{7} \label{eq:Dequiv2}
\KET{00} & \quad & -1 &
\stackrel{\VEC D}{\longleftrightarrow}~ & 
\UPS\SIG_\LAB{z}^{\,1}\UPS\SIG_\LAB{z}^{\,2} & \quad &
\UPS\SIG_\LAB{z}^{\,1} &
\stackrel{\VEC D}{\longleftrightarrow}~ &
\UPS\SIG_\LAB{z}^{\,2}
~\quad \nonumber \\
\KET{01} & \quad & \UPS\SIG_\LAB{y}^{\,2} &
\stackrel{\VEC D}{\longleftrightarrow}~ & 
\UPS\SIG_\LAB{z}^{\,1}\UPS\SIG_\LAB{x}^{\,2} & \quad &
-\UPS\SIG_\LAB{x}^{\,2} &
\stackrel{\VEC D}{\longleftrightarrow}~ &
\UPS\SIG_\LAB{z}^{\,1}\UPS\SIG_\LAB{y}^{\,2} \\ \nonumber
\KET{10} & \quad & \UPS\SIG_\LAB{y}^{\,1} &
\stackrel{\VEC D}{\longleftrightarrow}~ & 
\UPS\SIG_\LAB{x}^{\,1}\UPS\SIG_\LAB{z}^{\,2} & \quad &
-\UPS\SIG_\LAB{x}^{\,1} &
\stackrel{\VEC D}{\longleftrightarrow}~ &
\UPS\SIG_\LAB{y}^{\,1}\UPS\SIG_\LAB{z}^{\,2} \\ \nonumber
\KET{11} & \quad & -\UPS\SIG_\LAB{y}^{\,1}\UPS\SIG_\LAB{y}^{\,2} &
\stackrel{\VEC D}{\longleftrightarrow}~ &
\UPS\SIG_\LAB{x}^{\,1}\UPS\SIG_\LAB{x}^{\,2} & \quad &
\UPS\SIG_\LAB{x}^{\,1}\UPS\SIG_\LAB{y}^{\,2} &
\stackrel{\VEC D}{\longleftrightarrow}~ &
\UPS\SIG_\LAB{y}^{\,1}\UPS\SIG_\LAB{x}^{\,2}
\end{alignat}
(where the two columns differ by operation with $\VEC K$).
From this it may be seen that any ``spinor'' in
$(\ALG G_3^+)^{\otimes2}\!/\VEC D$ can be written as
\begin{equation} \begin{split}
\EMB\Psi ~=~ & \left( \DAB{10pt} \right. \!
(\alpha_0 + \beta_0 \VEC K) -
\UPS\SIG_\LAB{y}^{\,2} (\alpha_1 + \beta_1 \VEC K)
\, - \\ & \left. \DAB{10pt}
\UPS\SIG_\LAB{y}^{\,1} (\alpha_2 + \beta_2 \VEC K)
+ \UPS\SIG_\LAB{y}^{\,1} \UPS\SIG_\LAB{y}^{\,2}
(\alpha_3 + \beta_3 \VEC K) \right) \VEC D
\end{split} \end{equation}
(cf.\ \cite{SomLasDor:99}).
Alternatively, again using Eq.~(\ref{eq:Dequiv2}),
a unit norm spinor may be factorized into a product of
entities in the \emph{correlated and reduced} even subalgebra,
namely $\EMB\Psi ~=~ \VEC R^1\,\VEC S^2\,\VEC{TPDC}$,
where
\begin{equation} \begin{aligned}
\VEC R^1 ~\equiv~ & e^{-\UPS\phi\SIG_\LAB{z}^{\,1}/2} \,
e^{-\UPS\theta\SIG_\LAB{y}^{\,1}/2} ~, \\
\VEC S^2 ~\equiv~ & e^{-\UPS\varphi\SIG_\LAB{z}^{\,2}/2} \,
e^{-\UPS\vartheta\SIG_\LAB{y}^{\,2}/2} ~,
\end{aligned} \qquad \begin{aligned}
\VEC T ~\equiv~ & \cos(\varsigma/2) \,-\, \sin(\varsigma/2)\,
\SIG_\LAB{y}^{\,1}\SIG_\LAB{y}^{\,2}\VEC K ~, \\
\VEC P ~\equiv~ & e^{-\tau\VEC K/2} ~.
\end{aligned}
\end{equation}
Thus when $\varsigma = \pi$, the factor $\VEC T$ becomes
\begin{equation}
-\SIG_\LAB{y}^{\,1}\,\SIG_\LAB{y}^{\,2}\,\VEC K ~=~
(-\UPS\SIG_\LAB{y}^{\,1})(-\UPS\SIG_\LAB{y}^{\,2})\,\VEC K ~=~
e^{-(\pi/2)\UPS\SIG_\LAB{y}^{\,1}}\, e^{-(\pi/2)\UPS\SIG_\LAB{y}^{\,2}}\,
\VEC K ~,
\end{equation}
so that the arguments of the exponentials involving
$\UPS\SIG_\LAB{y}^{\,1}$ and $\UPS\SIG_\LAB{y}^{\,2}$
in the first two factors are shifted by $\pi/2$
while the total phase is shifted by $\tau = \pi$.
It follows that $\VEC T$ rotates the first two
factors in the planes defined by their conjugate
spinors $[-r_2^*; r_1^*], [-s_2^*; s_1^*]$.
Thus on right-multiplying by $\VEC E_{+}$ and expanding
in the usual basis, we obtain (up to an overall phase)
\renewcommand{\arraystretch}{1.4} 
\begin{equation} \begin{split} \label{eq:schmidt}
\UL{\EMB\Psi} ~=~ &
\cos(\TFRAC{\varsigma}2)\,e^{\imath\tau/2}
\begin{bmatrix}
\cos(\TFRAC{\theta}2)e^{\imath\phi/2} \\
\sin(\TFRAC{\theta}2)e^{-\imath\phi/2}
\end{bmatrix} \otimes \begin{bmatrix}
\cos(\TFRAC{\vartheta}2)e^{\imath\varphi/2} \\
\sin(\TFRAC{\vartheta}2)e^{-\imath\varphi/2}
\end{bmatrix}
\,+ \\ &
\sin(\TFRAC{\varsigma}2)\,e^{-\imath\tau/2}
\begin{bmatrix}
\sin(\TFRAC{\theta}2)e^{\imath\phi/2} \\
-\cos(\TFRAC{\theta}2)e^{-\imath\phi/2}
\end{bmatrix} \otimes \begin{bmatrix}
\sin(\TFRAC{\vartheta}2)e^{\imath\varphi/2} \\
-\cos(\TFRAC{\vartheta}2)e^{-\imath\varphi/2}
\end{bmatrix}
\end{split} \end{equation}
\renewcommand{\arraystretch}{1.0} 

This is known as the \emph{Schmidt decomposition} \cite{EkertKnigh:95}.
It is useful in studying the \emph{entanglement} of bipartite quantum systems,
which (in conventional terms) means that $\KET{\psi} \in \TEN H^{\otimes2}$
cannot be written as a product $\KET{\psi^1} \otimes \KET{\psi^2}
\equiv \KET{\psi^1}\KET{\psi^2} \equiv \KET{\psi^1\,\psi^2}$ for
any one-particle spinors $\KET{\psi^1}, \KET{\psi^2} \in \TEN H$.
In fact it is just the singular value decomposition in disguise,
since (for example) on arranging the entries of a two-qubit spinor
$\KET{\psi} = [ \psi_1 ;\cdots; \psi_4 ]$ in a $2\times2$ matrix,
we can write
\begin{equation}
\UL{\EMB\Psi} ~\equiv~ \begin{bmatrix}
\psi_1 & \psi_3 \\ \psi_2 & \psi_4 \end{bmatrix} ~=~
\UL{\VEC U\!} \, \UL{\VEC V\!} \, \UL{\VEC W\!}^{\,\dag}
~=~ \UL{\VEC u}^1 v^{11} {(\UL{\VEC w}^1)}^{\!\dag}
+ \UL{\VEC u}^2 v^{22} {(\UL{\VEC w}^2)}^{\!\dag} ~,
\end{equation}
where $\UL{\VEC V\!}\,$ is a $2\times2$ diagonal matrix
containing the singular values $v^{11} \ge v^{22} \ge0$
and $\UL{\VEC U\!}\,$, $\UL{\VEC W\!}\,$ are unitary matrices
with columns $\UL{\VEC u}^k$, $\UL{\VEC w}^k$, respectively.
Since the entries of the dyadic products
$\UL{\VEC u}^1{(\UL{\VEC w}^1)}^{\!\smash\dag}$,
$\UL{\VEC u}^2{(\UL{\VEC w}^2)}^{\!\smash\dag}$
are exactly the same as the Kronecker matrix products
$\UL{\VEC u}^1\otimes\UL{\VEC w}^1$,
$\UL{\VEC u}^2\otimes\UL{\VEC w}^2$,
the equivalence with Eq.\ (\ref{eq:schmidt}) follows with
$v^{11} \equiv \cos(\varsigma/2)$,
$v^{22} \equiv \sin(\varsigma/2)$,
and the Kronecker products of the columns
of $\UL{\VEC U\!}\,$ and $\UL{\VEC W\!}\,$ identified
with conjugate pairs of single qubit spinors whose
relative phases are given by $\exp(\pm\imath\tau/2)$.

Clearly a two-qubit spinor is unentangled
if and only if $v^{11} = 1$, which is equivalent
to $\varsigma = 0$ or $\VEC T = 1$.
Thus $\VEC T$ describes the entanglement of the
qubits, and is accordingly called the \emph{tangler}.
The geometric algebra approach clearly provides
deeper insight into the structure of entanglement
than does one based on mechanical matrix algebra.
In particular, the fact that $\tilde{\EMB\Psi}\EMB\Psi$
is even and reversion-symmetric in the Dirac
as well as the Pauli algebra implies that
it is the sum of a scalar and a four-vector
in the two-particle Dirac algebra $\ALG G_{2,6}\,$.
Since Lorentz transformations of the spinors cancel,
this entity is in fact a Lorentz invariant,
and dividing out the total phase $\VEC P$ as
$\VEC P(\tilde{\EMB\Psi}\EMB\Psi)\tilde{\VEC P}$
yields the \emph{square} of the tangler directly.
The availability of such powerful methods of
manipulating entities in the multiparticle Dirac
algebra promises to be useful in finding analogs of
the Schmidt decomposition for three or more qubits.

\section{Quantum Operations on Density Operators}
Quantum computers operate on information
stored in the states of quantum systems.
The systems are usually assumed to be arrays of
distinguishable qubits (two-state subsystems),
whose basis states $\KET0$ and $\KET1$ correspond
to the binary digits $0$ and $1$, respectively,
while the operations are usually taken to be unitary.
General unitary transformations of the qubits are built up
from simpler ones that affect only a few qubits at a time,
which are called \emph{quantum logic gates}.
The representation of these gates in suitable
products of Clifford algebras has been described
in Refs.~\cite{SomCorHav:98,Vlasov:01}.
The goal here will be to show how gates
act upon spinors in the even subalgebra,
and how they can be extended to a wider class of
nonunitary quantum operations on density operators.

Given the isomorphism between the algebra of
$2^N \times 2^N$ matrices over $\FLD C$ and
$\ALG G_3^{\otimes N} \!/ \VEC C$ relative
to a choice of basis in each particle space,
it is straightforward to interpret matrices in
the former as geometric entities in the latter.
A matrix $\UL{\VEC U\!} \in \GRP U(2^N)$,
however, does not generally correspond to an
entity $\VEC U$ in the \emph{even} subalgebra
${(\ALG G_3^+)}\DAB{1.6ex}^{\otimes\smash N} \!/ \VEC C$,
so that $\VEC U \EMB\Psi \not\in
{(\ALG G_3^+)}\DAB{1.6ex}^{\otimes\smash N} \!/ \VEC D$
for a general spinor
$\EMB\Psi \in {(\ALG G_3^+)}\DAB{1.6ex}^{\otimes\smash N} \!/ \VEC D$.
Nevertheless, letting $\VEC E_{-} \equiv \prod_q \VEC E_-^{\,q}$
be the idempotent ``opposite'' to $\VEC E_{+}$, and
noting that this satisfies $\VEC E_{+} \VEC E_{-} = 0$,
the product of $\VEC U\EMB\Psi$ with $\VEC E_{+}$ may be written as
\begin{equation}
\VEC U \EMB\Psi \VEC E_{+} ~=~
(\VEC U \EMB\Psi \VEC E_{+} \,+\,
\hat{\VEC U} \EMB\Psi \VEC E_{-}) \VEC E_{+} ~=~
2\, {\langle \VEC U \EMB\Psi \VEC E_{+} \rangle}_+ \, \VEC E_{+} ~,
\end{equation}
where the ``hat'' on $\hat{\VEC U}$ denotes
its image under inversion in the origin
(so that $\hat{\VEC E}_{+} = \VEC E_{-}$),
and hence ${\langle\,\UL{~}\,\rangle}_+$
is a projection onto the even subalgebra.
Because $({\ALG G_3^+)}\DAB{1.6ex}^{\otimes\smash N} \!/ \VEC C$
and $\GRP U(2^N)$ are both $(2^{2N})$-dimensional,
nothing is lost in this projection!
Thus we can drop the right-factor of $\VEC E_{+}$
as usual, and define the action of $\VEC U$ on $\EMB\Psi \in
{(\ALG G_3^+)}\DAB{1.6ex}^{\otimes\smash N} \!/ \VEC D$ as
\begin{equation} \label{eq:unit_act}
\VEC U \circ \VEC \Psi ~\equiv~
2\, {\langle \VEC U \EMB\Psi \VEC E_{+} \rangle}_+ ~.
\end{equation}
More generally, the usual action of the Pauli matrices
on spinors corresponds to the following action of the basis
vectors on the reduced even subalgebra \cite{DorLasGul:93}:
\begin{equation} \label{eq:pauli_rules}
\SIG_\mu \circ \EMB\Psi ~\equiv~ \SIG_\mu \EMB\Psi \SIG_\LAB{z} ~,
\quad \UPS \circ \EMB\Psi ~\equiv~ \UPS \EMB\Psi \SIG_\LAB{z}
\end{equation}

The simplest logic gate is the \textsf{NOT} of a single qubit,
which operates on the computational basis as follows:
\begin{equation}
\UL{\VEC N} \KET{0} ~=~ \KET{1} ~\leftrightarrow~ -\UPS\SIG_\LAB{y}
~,\quad
\UL{\VEC N} \KET{1} ~=~ \KET{0} ~\leftrightarrow~ 1
\end{equation}
Thus it might appear reasonable to represent the \textsf{NOT}
by $\VEC N \equiv \UPS\SIG_\LAB{y} \in \GRP{SU}(2)$, but when
$\UPS\SIG_\LAB{y}$ is applied a superposition $(1 - \UPS\SIG_\LAB{y})/
\sqrt2 \leftrightarrow (\KET{0} + \KET{1})/\sqrt2$, we get $(1 +
\UPS\SIG_\LAB{y})/\sqrt2 \leftrightarrow (\KET{0} - \KET{1})/\sqrt2$
instead of $(\KET0 + \KET1)/\sqrt2$ again.
For a single qubit this difference is just
an overall rotation by $\pi$ about $\SIG_\LAB{z}$,
but a second qubit can be affected by this phase
difference between the first qubit's states.
Therefore the correct representation of
the \textsf{NOT} gate in $\GRP{SU}(2)$ is
actually $\VEC N \equiv \pm\UPS\SIG_\LAB{x\,}$,
which preserves this superposition up
to an irrelevant overall phase shift:
$\imath\UL{\SIG}_\LAB{\,x} (\KET{0} +
\KET{1})/\sqrt2 ~\leftrightarrow$
\begin{equation} \begin{split}
& (\UPS\SIG_\LAB{x}) \circ (1 - \UPS\SIG_\LAB{y})/\sqrt2 ~=~
\UPS \circ \SIG_\LAB{x} \circ (1 - \UPS\SIG_\LAB{y})/\sqrt2 \\
=\;\; & \UPS \circ (\SIG_\LAB{x} (1 - \UPS\SIG_\LAB{y}) \SIG_\LAB{z} /\sqrt2)
~=~ \UPS \circ (-\UPS\SIG_\LAB{y} + 1)/\sqrt2 \\
=\;\; & (-\UPS\SIG_\LAB{y} + 1) \,\UPS\SIG_\LAB{z} /\sqrt2
~\leftrightarrow~ -\imath (\KET{0} + \KET{1}) / \sqrt2
\end{split} \end{equation}

More interesting logical operations on the
qubits must be able to transform the state
of one \emph{conditional} on that of another.
The usual way in which this is done is via the
\textsf{c-NOT} or \emph{controlled-NOT} gate.
As a matrix in $\GRP{SU}(4)$, this is
represented in the computational basis by
\begin{equation}
\UL{\VEC N}^{2|1} ~\equiv~ \sqrt{\imath}
\begin{bmatrix} 1&0&0&0 \\ 0&1&0&0 \\
0&0&0&1 \\ 0&0&1&0 \end{bmatrix} ~,
\end{equation}
which makes it clear that this operation \textsf{NOT}'s
the second qubit whenever the first is $1$.
The corresponding operator in geometric algebra is
\begin{equation}
\VEC N^{2|1} ~\equiv~ (1 + \UPS\SIG_\LAB{z}^{\,1})
/ \sqrt2 \, \left( \VEC E_+^{\,1} \,+\, \VEC E_-^{\,1}
\, \UPS\SIG_\LAB{x}^{\,2} \right) ~.
\end{equation}
This may also be written in exponential form as
\begin{equation}
e^{\,\UPS\pi\SIG_\LAB{z}^{\,1}/4} \,
e^{\,\UPS\pi\VEC E_-^{\,1}\SIG_\LAB{x}^{\,2}/2} ~=~
e^{-\UPS\pi (\VEC E_-^{\,1} (1 - \SIG_\LAB{x}^{\,2}) / 2 \,-\, 1/4)} ~.
\end{equation}
Physical implementations of this operation by
e.g.~NMR typically expand the exponential into
a product of relatively simple commuting factors
which can be performed sequentially \cite{HaSoTsCo:00}.

Note that since $(1 - \SIG_\LAB{x}^{\,2})/2$ is also an idempotent,
$\VEC N^{2|1}$ differs from $\VEC N^{1|2}$ by a swap
of the $\LAB{x}$ and $\LAB{z}$ axes for both qubits.
This self-inverse operation, called
the \emph{Hadamard transform} $\VEC H$,
is simply a rotation by $\pi$ about the
$(\SIG_\LAB{x} + \SIG_\LAB{z})/\sqrt2$ axis.
Sandwiching $\VEC N^{2|1}$ by Hadamards $\VEC H^{\,2} =
\UPS(\SIG_\LAB{x}^{\,2} + \SIG_\LAB{z}^{\,2}) / \sqrt2$
to just the second qubit gives
\begin{equation}
\VEC H^{\,2} \VEC N^{2|1} \VEC H^{\,2} ~=~
e^{-\UPS\pi (\VEC E_-^{\,1} \VEC H^{\,2} (1 -
\SIG_\LAB{x}^{\,2}) \VEC H^{\,2} / 2 - 1/4)} ~=~
e^{-\UPS\pi (\VEC E_-^{\,1} \VEC E_-^{\,2} \,-\, 1/4)} ~,
\end{equation}
so the \textsf{c-NOT} can also be viewed as a
rotated phase shift of the state $\KET{11}$ by $\pi$.
The Hadamard gate has the important feature of
transforming basis states into superpositions thereof;
indeed, as an element of the even subalgebra,
it actually represents the spinor of
a uniform superposition directly:
\begin{equation} \begin{split}
\UL{\VEC H} \KET{0} ~=~ &
\imath(\KET{0} + \KET{1})/\sqrt2 ~\leftrightarrow~
(\UPS\SIG_\LAB{z} + \UPS\SIG_\LAB{x})/\sqrt2 \, \VEC D
\\
\UL{\VEC H} \KET{1} ~=~ &
\imath(\KET{0} - \KET{1})/\sqrt2 ~\leftrightarrow~
(\UPS\SIG_\LAB{z} - \UPS\SIG_\LAB{x})/\sqrt2 \, \VEC D
\end{split} \end{equation}
Thus, by using the relations (\ref{eq:pauli_rules}),
we can show that applying a Hadamard to one of two qubits
in the state $\KET{11}$ followed by a \textsf{c-NOT}
gate to the other yields the entangled singlet state:
$\UL{\VEC N}^{2|1} \UL{\VEC H}^1 \KET{11} \leftrightarrow$
\begin{equation} \begin{split} \label{eq:singlet}
& \HALF \, \left( \DAB{12pt}
(1 + \UPS)\VEC E_+^1 + (1 - \UPS)\VEC E_-^1
\, \UPS\SIG_\LAB{x}^{\,2} \right) \circ \left(
(\UPS\SIG_\LAB{z}^{\,1} - \UPS\SIG_\LAB{x}^{\,1})
(-\UPS\SIG_\LAB{y}^{\,2})
\right) \\ =~ &
\HALF \, \left( \DAB{12pt}
\left( (1 + \UPS) \circ (\UPS\SIG_\LAB{z}^{\,1}) \right)
(-\UPS\SIG_\LAB{y}^{\,2}) +
\left( (1 - \UPS) \circ (-\UPS\SIG_\LAB{x}^{\,1}) \right)
\left( (\UPS\SIG_\LAB{x}^{\,2}) \circ (-\UPS\SIG_\LAB{y}^{\,2}) \right)
\right) \\ =~ &
\HALF (1 - \UPS\SIG_\LAB{z}^{\,1}) \UPS\SIG_\LAB{y}^{\,2} - \HALF
(\UPS\SIG_\LAB{x}^{\,1} - \UPS\SIG_\LAB{y}^{\,1}) \UPS\SIG_\LAB{z}^{\,2}
~\stackrel{\VEC D}{\longleftrightarrow}~
\TFRAC{\UPS}2 (\SIG_\LAB{y}^{\,2} + \SIG_\LAB{x}^{\,2}
- \SIG_\LAB{y}^{\,1} - \SIG_\LAB{x}^{\,1})
\end{split} \end{equation}
$\leftrightarrow \sqrt{-\imath}\, (\KET{10} - \KET{01})
/ \sqrt2 \equiv \sqrt{-\imath}\, \KET{\psi_{-\,}}$.
``Quantum'' gates like $\VEC H$ are not,
of course, found in conventional boolean logic,
and are an essential component of all quantum
algorithms that are more efficient than their
classical counterparts \cite{EkertJozsa:98,ClEkMaMo:98}.
Indeed, the \textsf{c-NOT} gate together with general
single qubit rotations are known to generate $\GRP{SU}(2^N)$,
and hence are \emph{universal} for quantum logic \cite{BBCDMSSSW:95}.

It turns out that unitary transformations
are not the most general sort of operation
that can be applied to a quantum system.
Most such \emph{quantum operations}, however,
produce a statistical outcome, and the ensemble of
possible outcomes must be described by a density operator.
The previous definition (Eq.~(\ref{eq:den_op_def}))
of the density operator of an ensemble of identical
and noninteracting qubits may be extended to an
ensemble of multi-qubit systems as follows:
\begin{equation}
\UL{\RHO} ~\equiv~ \OL{\KET{\psi}\BRA{\psi}} ~\leftrightarrow~
\RHO ~\equiv~ \OL{(\EMB\Psi \VEC D) \VEC E_{+} {(\EMB\Psi \VEC D)}^\sim}
~=~ \OL{\EMB\Psi \VEC E_{+} \tilde{\EMB\Psi}} \VEC C
\end{equation}
Suppressing the correlator $\VEC C$ as usual,
$\RHO$ may also be expressed in diagonal form as
\begin{equation}
\RHO ~=~ \VEC R \left( {\textstyle\sum}_{k=0}^{2^N}
\, \rho_k \VEC E_{(k)} \right) {\VEC R}^\dag ~=~
{\textstyle\sum}_{k=0}^{2^N} \, \rho_k \VEC r_k {\VEC r}_k^\dag ~.
\end{equation}
where $\VEC R \in \ALG G_3^{\otimes N} \!/ \VEC C$
corresponds to a unitary matrix $\UL{\VEC R} \in \GRP U(2^N)$
(in the usual $\SIG_\LAB{z}$ coordinate system), and
$0 \le \rho_k \le 1$ are the eigenvalues of $\RHO$.
The idempotents $\VEC E_{(k)}$ are given by $\prod_q
\VEC E\DAB{1.9ex}^{\,q}_{\smash{\epsilon_k^q}} \leftrightarrow
\KET{\chi_k^1\cdots\chi_k^N}\BRA{\chi_k^1\cdots\chi_k^N}$,
where $\epsilon_k^q \equiv 1-2\chi_k^q$ with
$\chi_k^q$ equal to the $q$-th bit in the binary
expansion of $k \in \{ 0, \ldots, 2^N-1 \}$.
It follows that $\KET{\rho_k} \leftrightarrow \VEC r_k \equiv
\VEC R \VEC E_{(k)}$ for $\rho_k > 0$ are the spinors of the
(unique, if $\rho_k \ne \rho_\ell~\forall k \ne l$)
minimal ensemble that realizes $\RHO$,
which therefore describes a pure state if
and only if it has rank $1$ as an operator.

Note that by Eq.~(\ref{eq:unit_act}), the density
operator transforms under unitary operations as
\begin{equation}
\RHO ~\mapsto~ \OL{(\VEC U \circ \EMB\Psi)
\VEC E_+ {(\VEC U \circ \EMB\Psi)}^\sim} ~=~
\VEC U\, \OL{\EMB\Psi \VEC E_+ \tilde{\EMB\Psi}} \,{\VEC U}^\dag
~=~ \VEC U \RHO\, {\VEC U}^\dag ~.
\end{equation}
Similarly, the ensemble-average expectation value of any observable
$\VEC O = {\VEC O}^\dag \in \ALG G_3^{\otimes N}\!/\VEC C$ is
\begin{equation}
\OL{\BRA{\psi}\,\UL{\VEC O}\,\KET{\psi}} ~\leftrightarrow~ 2^N \,
\OL{\langle \VEC E_+ \tilde{\EMB\Psi} \VEC O \EMB\Psi \VEC E_+ \rangle}_0
~=~ 2^N \, {\langle \VEC O\, \OL{\EMB\Psi \VEC E_+ \tilde{\EMB\Psi}} \rangle}_0
~\equiv~ 2^{N\,} {\langle \VEC O \RHO \, \rangle}_0 ~,
\end{equation}
just as shown in Eq.~(\ref{eq:rho1_obs}) for single qubit ensembles.
In contrast to the case of a single qubit, however, the geometric
interpretation of these observables is not straightforward.
While one can certainly express $\RHO$ as a finite ensemble
average $\sum_k p_k \EMB\Psi_k \VEC E_+ \tilde{\EMB\Psi}_{k\,}$
(where the $p_k > 0$ are probabilities with $\sum_k p_k = 1$),
this decomposition is highly nonunique.
The minimal ensemble obtained by diagonalization,
on the other hand, will generally include entangled
spinors $\VEC r_{k\,}$, for which the expectation value
${\langle \VEC O\, \VEC r_k^{\,} {\VEC r}_k^\dag \rangle}_0$
cannot be expressed as a product of inner products
of the factors of $\VEC O = \VEC O^1 \cdots \VEC O^N$
with the polarization vectors of the individual qubits
(indeed, $\VEC O$ itself need not be factorizable!).

The best one can do is to expand $\RHO$ in the
\emph{product operator} basis consisting of all
$2^{2N}$ products of the basis vectors $\SIG_\mu^{\,q}$, i.e.
\begin{equation}
\RHO ~=~ \sum_{ \mu^1,\ldots,\,\mu^N \in
\{ 0, \LAB{x}, \LAB{y}, \LAB{z} \} } \,
\rho_{\mu^1\cdots\mu^N} \,
\SIG_{\mu^1}^{\,1} \cdots \SIG_{\mu^N}^{\,N} ~,
\end{equation}
where $\rho_{\mu^1\cdots\mu^N} \in \FLD R$ and
$\SIG_0^{\,q} \equiv 1$ for notational convenience.
The utility of this basis is most simply demonstrated
via a concrete example, namely NMR spectroscopy.
Here one is given a liquid sample consisting of identical
molecules whose nuclear spins are chemically distinguishable,
and hence constitutes an ensemble of multi-qubit systems (see
\cite{CorPriHav:98,HaCoSoTs:00,HaSoTsCo:00} and references therein).
The energy of interaction between the spins and an
external magnetic field along $\LAB{z}$ is given
by an observable called the Zeeman Hamiltonian,
$\VEC Z \equiv (\omega^1 \SIG_\LAB{z}^{\,1}
+\cdots+ \omega^N \SIG_\LAB{z}^{\,N})/2$,
where $\omega^q$ is the energy difference between the
$\KET0$ and $\KET1$ states of the $q$-th spin in the field.
In thermal equilibrium at room temperatures,
the polarization of the spins relative to the strongest
available fields is typically $\alpha \sim 10^{-6}$,
and the density operator of the ensemble is essentially $\RHO_\LAB{eq}
= 2^{-N}(1 + \alpha(\SIG_\LAB{z}^{\,1} +\cdots+ \SIG_\LAB{z}^{\,N}))$.
Via a suitable pulse of radio-frequency radiation,
this may be rotated to $\RHO_\LAB{eq}' \equiv 2^{-N}(1 +
\alpha(\SIG_\LAB{x}^{\,1} +\cdots+ \SIG_\LAB{x}^{\,N}))$,
which evolves under the interaction with the field as
\begin{equation}
e^{-\UPS\VEC Z t} \RHO_\LAB{eq}' e^{\UPS\VEC Z t} ~=~
\begin{aligned}[t] & 2^{-N}\left( 1 + \alpha( \cos(\omega^1t)
\SIG_\LAB{x}^{\,1} - \sin(\omega^1t) \SIG_\LAB{y}^{\,1} \,+\cdots
\right. \\ & \left.
\cdots+\, \cos(\omega^Nt) \SIG_\LAB{x}^{\,N} -
\sin(\omega^Nt) \SIG_\LAB{y}^{\,N}) \right) ~.
\end{aligned}
\end{equation}
Thus on measuring the total magnetization
$M_\LAB{x}$ along the $\LAB{x}$ axis, $\VEC O \equiv
\gamma (\SIG_\LAB{x}^{\,1} +\cdots+ \SIG_\LAB{x}^{\,N})$
(where $\gamma$ is the nuclear gyromagnetic ratio),
one obtains the sum of the projections of the rotating
magnetization vectors of the spins along the $\LAB{x}$-axis, i.e.
\begin{equation}
M_\LAB{x}(t) ~=~ \alpha\gamma(\cos(\omega^1t) +\cdots+ \cos(\omega^Nt)) ~,
\end{equation}
whose Fourier transform reveals the contribution from each spin.
The way in which the factors of product operators transform
like vectors under rotations accounts in large part for
the computational utility of the product operator basis.
Of course, unless it is a natural part of the problem at hand
(as in NMR), one is better off not chosing a basis at all!

A \emph{normal quantum operation} is a linear
transformation of the density operator that may be
written in operator sum form as \cite{Kraus:83}
\begin{equation}
\RHO ~\mapsto~ \Omega(\RHO) ~\equiv~ {\textstyle\sum}_k
\, \VEC Q_k \, \RHO \, {\VEC Q}_k^\dag ~,
\end{equation}
where the \emph{Kraus operators}
$\VEC Q_k \in \ALG G_3^{\otimes N}\!/\VEC C$
satisfy $\sum_k {\VEC Q}_k^\dag \VEC Q_k = 1$.
The term ``normal'' here\footnote{
We prefer to avoid the more common but clumsy
and matrix-bound term ``trace-preserving''.}
refers to the fact that such an operation
preserves the scalar part of $\RHO$, since
\begin{equation}
\left\langle \Omega(\RHO) \right\rangle_0 ~=~
{\textstyle\sum}_k \left\langle \, \VEC Q_k \,
\RHO \, {\VEC Q}_k^\dag \right\rangle_0 ~=~
\left\langle \, \RHO \, {\textstyle\sum}_k \,
{\VEC Q}_k^\dag \VEC Q_k \, \right\rangle_0 ~=~ 2^{-N} ~.
\end{equation}
It is also easily seen that such quantum
operations are \emph{positive}, in that they
preserve the positive-definiteness of $\RHO$;
in fact, these operations have a yet stronger
property known as \emph{complete positivity},
meaning that if the qubits to which $\Omega$
applies are embedded in a larger system,
then applying $\Omega$ to just those qubits preserves the
positive-definiteness of the larger system's density operator.
That this is a nontrivial extension of positivity is shown by
the two-qubit ``partial transpose'' operator $\Omega_\LAB{T}^1$,
which carries $\SIG_\LAB{y}^{\,1} \mapsto -\SIG_\LAB{y}^{\,1}$
but leaves all the other operator factors unchanged;
this is clearly positive on density operators not
involving the second qubit, but acts on the density
operator of the singlet state (Eq.~(\ref{eq:singlet})) as
\begin{equation} \begin{split}
& \EMB\psi_- \begin{aligned}[t] \equiv~ & \TFRAC14 (\SIG_\LAB{x}^{\,1}
+ \SIG_\LAB{y}^{\,1} - \SIG_\LAB{x}^{\,2} - \SIG_\LAB{y}^{\,2})
\VEC E_+^1 \VEC E_+^2 (\SIG_\LAB{x}^{\,1} + \SIG_\LAB{y}^{\,1}
- \SIG_\LAB{x}^{\,2} - \SIG_\LAB{y}^{\,2}) \\ =~ &
\TFRAC14 (1 - \SIG_\LAB{x}^{\,1} \SIG_\LAB{x}^{\,2}
- \SIG_\LAB{y}^{\,1} \SIG_\LAB{y}^{\,2}
- \SIG_\LAB{z}^{\,1} \SIG_\LAB{z}^{\,2})
\end{aligned} \\ \mapsto\quad &
\Omega_\LAB{T}^1(\EMB\psi_-) ~=~
\TFRAC14 (1 - \SIG_\LAB{x}^{\,1} \SIG_\LAB{x}^{\,2}
+ \SIG_\LAB{y}^{\,1} \SIG_\LAB{y}^{\,2}
- \SIG_\LAB{z}^{\,1} \SIG_\LAB{z}^{\,2}) ~,
\end{split} \end{equation}
which has eigenvalues $[1/2,\,1/2,\,1/2,\,-1/2]$.

A quantum operation $\Omega$ is called \emph{unital} if
it preserves the identity itself, i.e. $\Omega(1) = 1$,
or equivalently, $\sum_k \VEC Q_k {\VEC Q}_k^\dag = 1$.
Perhaps the most important example of a normal
unital operation is found in the \emph{contraction}%
\footnote{Otherwise known as the ``partial trace''.}
by a single qubit $q \in \{1,\ldots,N\}$, which may
be written in operator sum form as \cite{SomCorHav:98}:
\begin{equation}
2\, \left\langle \, \RHO \, \right\rangle^q ~\equiv~
\VEC E_+^q \RHO \VEC E_+^q + \VEC E_-^q \RHO \VEC E_-^q
+ \SIG_\LAB{x}^{\,q} (\VEC E_+^q \RHO \VEC E_+^q +
\VEC E_-^q \RHO \VEC E_-^q) \SIG_\LAB{x}^{\,q}
\end{equation}
This may also be expressed by dropping all terms in the
product operator expansion of $\RHO$ depending on $q$,
and multiplying the remaining terms by a factor of $2$.
Note that, while ${\langle\,\UL{~}\,\rangle}^q$ is normal and unital,
this factor means that the contraction itself is neither.
The factor is nevertheless required if the result is to be
interpreted as a density operator for the remaining qubits,
since the contraction by the second qubit
of the above singlet state is ${\langle \,
\EMB\psi_- \, \rangle}^2 ~=~ 1/4$ (not $1/2$).

This example also illustrates an important way in which
general quantum operations are realized in practice,
despite the fact that the universe as a whole evolves unitarily.
As shown previously, the superposition state with spinor
$\EMB\Psi^1 = (1 - \UPS\SIG_\LAB{y}^{\,1})/\sqrt2$ is
converted into the singlet state with density operator
$\EMB\psi_-$ by letting it interact with a second qubit
so as to effect the \textsf{c-NOT} operation $\VEC N^{2|1}$.
The contraction then corresponds to ``discarding'' the second qubit
(i.e.~ensuring that it does not further interact with the first
and hence can be ignored), which yields the density operator
$1/2$ of the totally mixed state for the first qubit.
Since the basis states $\VEC E_\pm^{\,1}$ are unaffected
by $\VEC N^{2|1}$, the net quantum operation on the
first qubit corresponds to what is known in quantum
communications theory as the \emph{phase damping channel}
\begin{equation}
\RHO ~\mapsto~ (1 - p) \RHO \,+\, p\, \VEC E_+
\RHO \VEC E_+ \,+\, p\, \VEC E_- \RHO \VEC E_-
\end{equation}
with damping parameter $p = 1$.
Phase damping is also known as $T_2$ relaxation in NMR,
and as \emph{decoherence} in quantum information processing;
it is widely believed to be the dominant mechanism by
which classical statistical mechanics arises from the
underlying unitary dynamics \cite{GiuliniEtAl:96}.

To illustrate the utility of geometric algebra
in the study of general quantum operations,
an eigenvalue characterization of normal, unital,
one-bit quantum operations $\Omega$ will now be derived.
This characterization was originally given
by Fujiwara \& Algoet \cite{FujiwAlgoe:99},
although the derivation here parallels that more recently
obtained using matrix methods King \& Ruskai \cite{KingRuskai:00}.
This derivation will regard the Kraus operators $\VEC Q_k
\in \ALG G_3$ as ``complex quaternions'' $\VEC A_k +
\UPS \VEC B_k$ with $\VEC A_k, \VEC B_k \in \ALG G_3^+$,
and consider the action of an arbitrary operation
$\Omega$ on the scalar and vector parts of
$\RHO \equiv (1 + \VEC r) / 2$ separately.

First, the action on $1$ is
\begin{equation} \begin{split}
\Omega(1) ~=~ &
{\textstyle\sum}_k \, (\VEC A_k + \UPS \VEC B_k)
{(\VEC A_k + \UPS \VEC B_k)}^\dag
\\ =~ & {\textstyle\sum}_k
\left( \VEC A_k \tilde{\VEC A}_k + \VEC B_k \tilde{\VEC B}_k \right)
+ \UPS\, {\textstyle\sum}_k
\left( \VEC B_k \tilde{\VEC A}_k - \VEC A_k \tilde{\VEC B}_k ) \right) ~.
\end{split} \end{equation}
The first summation is symmetric with respect
to spatial reversion and inversion, i.e.~scalar,
while the second (excluding the $\UPS$) is reversion
antisymmetric but inversion symmetric, i.e.~a bivector.
Writing $\VEC A \equiv \alpha + \UPS \VEC a$
and $\VEC B \equiv \beta + \UPS \VEC b$,
so that $\VEC A_k \tilde{\VEC A}_k + \VEC B_k \tilde{\VEC B}_k
= \alpha_k^2 + \| \VEC a_k \|^2 + \beta_k^2 + \| \VEC b_k \|^2$,
we may further expand the second summation as follows:
\begin{equation} \begin{split}
& {\textstyle\sum}_k \left[ \,
(\beta_k + \UPS\VEC b_k)(\alpha_k - \UPS\VEC a_k) -
(\alpha_k + \UPS\VEC a_k)(\beta_k - \UPS\VEC b_k) \,
\right] \\ =~ &
{\textstyle\sum}_k \begin{aligned}[t] &
\left[ \left( \alpha_k\beta_k + \UPS (\alpha_k \VEC b_k
- \beta_k \VEC a_k) + \VEC b_k \VEC a_k \right)
\DAB{10pt} \right. \, - \\ & \left. \DAB{10pt}
\left( \alpha_k\beta_k + \UPS (\beta_k \VEC a_k
- \alpha_k \VEC b_k) + \VEC a_k \VEC b_k \right)
\right] \end{aligned} \\ =~ &
{\textstyle\sum}_k \left[\, 2 \UPS ( \alpha_k \VEC b_k -
\beta_k \VEC a_k ) - 2 \VEC a_k \wedge \VEC b_k \, \right]
\end{split} \end{equation}
Thus $\Omega$ is unital if and only if
\begin{equation}
{\textstyle\sum}_k \left( \alpha_k \VEC b_k -
\beta_k \VEC a_k \right) ~=~ {\textstyle\sum}_k
\, \VEC a_k \times \VEC b_k ~, \label{eq:cond1a}
\end{equation}
(where ``$\times$'' denotes the cross product) and
\begin{equation}
{\textstyle\sum}_k \left(
\alpha_k^2 + \| \VEC a_k \|^2 + \beta_k^2 + \| \VEC b_k \|^2
\right) ~=~ 1 ~. \label{eq:cond1b}
\end{equation}

Similarly, the action on $\VEC r \in \TEN R_3$ is
\begin{equation} \begin{split}
\Omega(\VEC r) ~=~ &
{\textstyle\sum}_k \, (\VEC A_k + \UPS \VEC B_k) \,
\VEC r \, (\tilde{\VEC A}_k - \UPS \tilde{\VEC B}_k)
\\ =~ & {\textstyle\sum}_k \left(
\VEC A_k \VEC r \tilde{\VEC A}_k + \VEC B_k \VEC r \tilde{\VEC B}_k
\right) + \UPS\, {\textstyle\sum}_k \left(
\VEC B_k \VEC r \tilde{\VEC A}_k - \VEC A_k \VEC r \tilde{\VEC B}_k
\right) ~.
\end{split} \end{equation}
The first summation is over different dilation/rotations of $\VEC r$;
the second summation (excluding the $\UPS$) is reversion and inversion
antisymmetric, i.e.~a trivector, and may be further expanded as above:
\begin{equation} \begin{split}
& {\textstyle\sum}_k \left[ \,
(\beta_k + \UPS\VEC b_k) \, \VEC r \, (\alpha_k - \UPS\VEC a_k) -
(\alpha_k + \UPS\VEC a_k) \, \VEC r \, (\beta_k - \UPS\VEC b_k) \,
\right] \\ =~ &
{\textstyle\sum}_k \begin{aligned}[t] &
\left[ \left( \alpha_k\beta_k\VEC r + \UPS (\alpha_k \VEC b_k \VEC r
- \beta_k \VEC r \VEC a_k) + \VEC b_k \VEC r \VEC a_k \right)
\DAB{10pt} \right. \, - \\ & \left. \DAB{10pt}
\left( \alpha_k\beta_k\VEC r + \UPS (\beta_k \VEC a_k \VEC r
- \alpha_k \VEC r \VEC b_k) + \VEC a_k \VEC r \VEC b_k \right)
\right] \end{aligned} \\ =~ &
{\textstyle\sum}_k \left[
\, 2\UPS \alpha_k \VEC b_k \cdot \VEC r
-  2\UPS \beta_k  \VEC a_k \cdot \VEC r
+ 2 \VEC b_k \wedge \VEC r \wedge \VEC a_k
\, \right]
\end{split} \end{equation}
Multiplying through by $-\UPS/2$ converts this to
\begin{equation}
{\textstyle\sum}_k \left[
\, \alpha_k \VEC b_k \cdot \VEC r - \beta_k  \VEC a_k \cdot \VEC r
- (\UPS (\VEC a_k \wedge \VEC b_k)) \cdot \VEC r
\, \right] ~,
\end{equation}
which vanishes if and only if
\begin{equation}
{\textstyle\sum}_k \left( \, \alpha_k \VEC b_k
- \beta_k  \VEC a_k \, \right) \cdot \VEC r ~=~
{\textstyle\sum}_k \, (\VEC b_k \times \VEC a_k) \cdot \VEC r ~.
\end{equation}
If $\Omega$ is normal, this must be true
for all $\VEC r$, which is equivalent to
\begin{equation}
{\textstyle\sum}_k \left( \, \label{eq:cond2}
\alpha_k \VEC b_k - \beta_k  \VEC a_k \, \right)
~=~ {\textstyle\sum}_k \, \VEC b_k \times \VEC a_k ~.
\end{equation}
A comparison with Eq.~(\ref{eq:cond1a}) shows further
that $\Omega$ is both unital and normal if and only
if $\sum_k \alpha_k \VEC b_k = \sum_k \beta_k \VEC a_k$
and $\sum_k \VEC a_k \times \VEC b_k = 0$.

If we regard a normal $\Omega$ as an affine transformation, i.e.
\begin{equation}
\Omega( \HALF ( 1 + \VEC r ) ) ~=~
\HALF ( 1 + \VEC t + {\textstyle\sum}_\mu \,
(\SIG_\mu \cdot \VEC r ) \VEC s_\mu )
\end{equation}
($\VEC s_\mu, \VEC t \in \TEN R_3$),
we see from the derivation leading up to (\ref{eq:cond1a}) that
\begin{equation} \begin{split}
\VEC t ~=~ & 2 \, {\textstyle\sum}_k \left( \alpha_k \VEC b_k
- \beta_k \VEC a_k - \VEC a_k \times \VEC b_k \right)
\\ =~ & 4 \, {\textstyle\sum}_k
\left( \alpha_k \VEC b_k - \beta_k \VEC a_k \right) ~=~
4 \, {\textstyle\sum}_k \, \VEC b_k \times \VEC a_k
\quad (\text{by (\ref{eq:cond2}))} ~.
\end{split} \end{equation}
Similarly, the above vectors $\VEC s_\mu =
{\left\langle\Omega(\SIG_\mu)\right\rangle}_1$ are
\begin{equation}
\VEC s_\mu ~=~ {\textstyle\sum}_k \left(
\VEC A_k \SIG_\mu \tilde{\VEC A}_k +
\VEC B_k \SIG_\mu \tilde{\VEC B}_k \right)
\quad (\mu \in \{\LAB{x},\LAB{y},\LAB{z}\}) ~,
\end{equation}
i.e.~a sum of independent dilation/rotations of each basis vector.

It follows that normal and unital quantum operations
$\Omega$ may be characterized by finding conditions
for the linear map $\VEC r \mapsto {\left\langle\Omega
(\VEC r)\right\rangle}_1 = \Omega(\VEC r)$ to be
written as a sum of dilation/rotations of $\VEC r$.
To this end, we expand $\VEC A_k \VEC r \tilde{\VEC A}_k$ as
\begin{equation} \begin{split}
(\alpha_k + \UPS \VEC a_k) \, \VEC r \, (\alpha_k - \UPS \VEC a_k)
~=~ & \alpha_k^2 \VEC r + \alpha_k \UPS (\VEC a_k\VEC r
- \VEC r \VEC a_k) + \VEC a_k \VEC r \VEC a_k \\
=~ & \alpha_k^2 \VEC r + 2 \alpha_k \VEC r \times \VEC a_k + 2 (\VEC r
\cdot \VEC a_k) \VEC a_k - \| \VEC a_k \|^2 \VEC r ~,
\end{split} \end{equation}
with a similar expansion for $\VEC B_k \VEC r \tilde{\VEC B}_{k\,}$.
Thus on assuming that $\Omega$ is diagonal, i.e.~$\VEC s_\mu
\equiv \lambda_\mu \SIG_\mu = \Omega( \SIG_\mu )$, we get
\begin{equation}
\lambda_\mu \SIG_\mu ~=~ \begin{aligned}[t] &
\SIG_\mu \, {\textstyle\sum}_k (\alpha_k^2 + \beta_k^2
- \| \VEC a_k \|^2 - \| \VEC b_k \|^2) + 2\, \SIG_\mu \times
{\textstyle\sum}_k (\alpha_k \VEC a_k + \beta_k \VEC b_k)
\\ & +\, 2\,
{\textstyle\sum}_k \left( (\SIG_\mu \cdot \VEC a_k) \VEC a_k
+ (\SIG_\mu \cdot \VEC b_k) \VEC b_k \right) ~. \end{aligned}
\end{equation}
Dotting both sides by $\SIG_\mu$ now yields
\begin{equation} \begin{split}
\lambda_\mu ~=~ & {\textstyle\sum}_k \left(
\alpha_k^2 + \beta_k^2 - \| \VEC a_k \|^2 - \| \VEC b_k \|^2 +
2 \, (\SIG_\mu \cdot \VEC a_k)^2 + 2 \, (\SIG_\mu \cdot \VEC b_k)^2
\right) \\ =~ & 1 - 2\, {\textstyle\sum}_k \left(
\| \VEC a_k \|^2 + \| \VEC b_k \|^2 - (\SIG_\mu
\cdot \VEC a_k)^2 - (\SIG_\mu \cdot \VEC b_k)^2
\right) \\ =~ & 1 + 2\, {\textstyle\sum}_k \left(
(\SIG_\mu \wedge \VEC a_k)^2 + (\SIG_\mu \wedge \VEC b_k)^2
\right) ~,
\end{split} \end{equation}
so we have simple expressions for the eigenvalues.
Now consider the vector obtained from the first line of this equation,
i.e.~$\sum_{\mu\in\{\LAB{x},\LAB{y},\LAB{z}\}} \, \lambda_\mu \SIG_\mu$
\begin{equation} \begin{split}
=~ & {\textstyle\sum}_\mu \, \SIG_\mu\, {\textstyle\sum}_k \left(
\alpha_k^2 + \beta_k^2 - \| \VEC a_k \|^2 - \| \VEC b_k \|^2 +
2\,(\SIG_\mu \cdot \VEC a_k)^2 + 2\,(\SIG_\mu \cdot \VEC b_k)^2 \right)
\\ =~ &
\VEC p_0\; {\textstyle\sum}_k \left( \alpha_k^2 + \beta_k^2
\right) + {\textstyle\sum}_\mu \, \VEC p_\mu \;
{\textstyle\sum}_k \left( (\SIG_\mu \cdot \VEC a_k)^2
+ (\SIG_\mu \cdot \VEC b_k)^2 \right) ~,
\end{split} \end{equation}
where
\begin{alignat}{2}
\VEC p_0 ~\equiv~ & ~\SIG_\LAB{x} + \SIG_\LAB{y} + \SIG_\LAB{z}~, \quad &
\VEC p_\LAB{x} ~\equiv~ & ~\SIG_\LAB{x} - \SIG_\LAB{y} - \SIG_\LAB{z}~, \\
\VEC p_\LAB{y} ~\equiv~ & -\SIG_\LAB{x} + \SIG_\LAB{y} - \SIG_\LAB{z}~, \quad &
\VEC p_\LAB{z} ~\equiv~ & -\SIG_\LAB{x} - \SIG_\LAB{y} + \SIG_\LAB{z}~.
\nonumber
\end{alignat}
Since the coefficients of the $\VEC p$'s are nonnegative
and sum to $1$ by (\ref{eq:cond1b}), this shows
that the vector $\sum_\mu \lambda_\mu \SIG_\mu$
lies within the tetrahedron $\langle \VEC p_0,
\VEC p_\LAB{x}, \VEC p_\LAB{y}, \VEC p_\LAB{z} \rangle$,
which is the condition on the eigenvalues found
by Fujiwara and Algoet \cite{FujiwAlgoe:99} as
well as by King and Ruskai \cite{KingRuskai:00}.

It is also known that an arbitrary linear map
$\Omega$ has an operator sum representation if and
only if it is completely positive \cite{Schumacher:96},
so the above can also be viewed as a characterization
of complete positivity for normal and unital
maps of a single qubit's density operator.
Finally, it is worth stressing once again that,
because of the isomorphisms which exist between the
Pauli algebra and the even subalgebra of the Dirac algebra,
every step of the above derivation carries with it a natural
interpretation in space-time, and is in fact even easier to
carry out when the full power of the Dirac algebra is used.

In conclusion, it is hoped that the forgoing
has given the reader a taste of the new insights
which geometric algebra can provide into quantum
information processing --- and an appetite for more!

\newcommand{\etalchar}[1]{$^{#1}$}
\providecommand{\bysame}{\leavevmode\hbox to3em{\hrulefill}\thinspace}

\end{document}